\documentclass{aa}  

\usepackage{ulem}
\usepackage{graphicx}
\usepackage{xcolor}
\usepackage{booktabs}
\usepackage{adjustbox}
\usepackage{placeins}
\usepackage{stfloats}
\usepackage{rotating}
\usepackage{txfonts}
\newcommand{\rgc}{$R_\mathrm{GC}$}

\begin{document}

   \title{JOYS: The [D/H] abundance derived from protostellar outflows across the Galactic disk measured with JWST}
   \titlerunning{[D/H] in outflows across the Galactic disk}

   \author{
% core JOYS members
L.~Francis\inst{1} \and
E.~F.~van~Dishoeck\inst{1, 3} \and
A.~Caratti~o~Garatti\inst{6} \and
M.~L.~van~Gelder\inst{1} \and
C.~Gieser\inst{3, 2} \and
H.~Beuther\inst{2} \and
T.P.~Ray\inst{10}
L.~Tychoniec\inst{1} \and
P.~Nazari \inst{4}
S.~Reyes\inst{2}
%G.~Perotti\inst{2} \and
P.~J. Kavanagh\inst{5} \and
%T.~Ray\inst{7} \and
P.~Klaassen\inst{7} \and
% K.~Justtanont\inst{9} \and
% H.~Linnartz\inst{10} \and
% W.~R. M. Rocha\inst{1,10} \and
% K.~Slavicinska\inst{1, 10} \and
% MIRI PIs
%L.~Colina\inst{11} \and
%T.~Greve\inst{12} \and
M.~Güdel\inst{8, 2, 9} \and
T.~Henning\inst{2}
%P.-O.~Lagage\inst{13} \and
%G.~Östlin\inst{14}
%B.~Vandenbussche\inst{17} \and
%C.~Waelkens\inst{17} \and
%G.~Wright\inst{8}
          }     

\institute{
Leiden Observatory, Leiden University, PO Box 9513, 2300 RA Leiden, The Netherlands \and
Max Planck Institute for Extraterrestrial Physics, Gießenbachstraße 1, 85749 Garching bei München \and
Max Planck Institute for Astronomy, Königstuhl 17, 69117 Heidelberg, Germany \and
European Southern Observatory, Karl-Schwarzschild-Strasse 2, 85748 Garching bei München, Germany \and
Department of Experimental Physics, Maynooth University-National University of Ireland Maynooth, Maynooth, Co Kildare, Ireland \and
INAF-Osservatorio Astronomico di Capodimonte, Salita Moiariello 16, I-80131 Napoli, Italy \and
UK Astronomy Technology Centre, Royal Observatory Edinburgh, Blackford Hill, Edinburgh EH9 3HJ, UK \and
Department of Astrophysics, University of Vienna, Türkenschanzstr. 17, 1180 Vienna, Austria \and
ETH Zürich, Institute for Particle Physics and Astrophysics, Wolfgang-Pauli-Str. 27, 8093 Zürich, Switzerland
School of Cosmic Physics, Dublin Institute for Advanced Studies, 31 Fitzwilliam Place, Dublin 2, Ireland
}

   \date{}

% \abstract{}{}{}{}{} 
% 5 {} token are mandatory

  \abstract
    % context heading (optional), leave it empty if necessary
   {The total deuterium abundance [D/H] in the universe is set by just two processes: the creation of deuterium in Big Bang Nucleosynthesis at an abundance of [D/H]$=2.58\pm0.13\times10^{-5}$, and its destruction within stellar interiors (astration). Measurements of variations in the total [D/H] abundance can thus potentially provide a probe of Galactic chemical evolution. However, most observational measurements of [D/H] are only sensitive to the gas-phase deuterium, and the amount of deuterium sequestered in dust grains is debated. With the launch of the James Webb Space Telescope (JWST), it is now possible to measure the gas-phase [D/H] at unprecedented sensitivity and distances through observation of mid-IR lines of H$_2$ and HD. Comparisons of gas-phase [D/H] with the constraints on the total [D/H] from the primordial abundance and Galactic chemical evolution models can provide insight into the degree of Deuterium lock-up in grains and the star formation history of our Galaxy.}
   {We use data from the JWST Observations of Young protoStars (JOYS) program of 5 nearby and resolved low-mass protostellar outflows and 5 distant high-mass protostellar outflows taken with the JWST Mid Infrafred Instrument (MIRI) Medium Resolution Spectrometer (MRS) to measure gas-phase [D/H] via H$_2$ and HD lines, assuming the gas is fully molecular.} 
   {We extract spectra from various locations in the outflows. Using a rotational diagram analysis covering lines of H$_2$ and HD with similar excitation energies, we derive the column density of HD and H$_2$ or their upper limits. We then calculate the gas-phase [D/H] from the column density results, and additionally apply a correction factor for the effect of chemical conversion of HD to atomic D and non-LTE excitation on the HD abundance in the shocks. To investigate the spatial distribution of the bulk gas and species refractory species associated with the dust grains, we also construct integrated line intensity maps of H$_2$, HD, [Fe II], [Fe I], and [S I] lines.}
   {A comparison of gas-phase [D/H] between our low-mass sources shows variations of up to a factor of $\sim4$, despite these sources likely having formed in nearly the same region of the Galactic disk that would be expected to have nearly constant total [D/H]. Most measurements of gas-phase [D/H] from our work or previous studies produce [D/H] $\lesssim 1.0\times10^{-5}$, a factor of $2-4$ lower than found from local UV absorption lines and as expected from Galactic chemical evolution models. In the integrated line intensity maps, the morphology of the HD R(6) line emission is strongly correlated with the H$_2$ S(7), [S I], and [Fe I] lines which mostly trace high velocity jet knots and bright bow-shocks. In our extracted spectra along the outflows, there is similarly a strong correlation between the H$_2$ and HD column density and the [S I] and [Fe I] line flux, however, no correlation is seen between [D/H] and the [S I] or [Fe I] line flux.}
   {The variations in [D/H] between our low-mass sources and the low [D/H] with respect to Galactic chemical evolution models suggest that our observations are not sensitive to the total [D/H]. Significant depletion of deuterium onto carbonaceous dust grains is a possible explanation, and tentative evidence of enhanced [D/H] towards positions with higher gas-phase Fe abundance is seen in the HH 211 outflow. Deeper observations of HD and H$_2$ across a wider range of shock conditions and modelling of the carbonaceous dust-grain destruction and shock conditions are warranted to test for the effects of depletion.}
   
   \keywords{astrochemistry -- stars -- Galaxy: abundances -- Infrared: ISM -- ISM: jets and outflows -- Stars: formation
               }

   \maketitle
%
%-------------------------------------------------------------------

\section{Introduction}
\label{sec:intro}

% origin of deuterium
Deuterium is expected to only be created during Big Bang Nucleosynthesis (BBN). Afterwards, the deuterium abundance in the universe is depleted as it is incorporated into stellar interiors and burned into $^3$He in the process of astration \citep{Epstein1976}. There is no known nucleosynthetic process to replenish the supply of deuterium, and thus the deuterium abundance with respect to hydrogen ([D/H]) has been proposed as a robust measure of stellar chemical evolution; with more generations of star formation, there should be an increasing destruction of deuterium by astration \citep{Sarkar1996}. The primordial [D/H] set by BBN has been accurately determined from UV observations of ultraviolet D I and H I absorption lines in high-redshift, metal-poor quasars to be $2.53 \pm 0.04 \times 10^{-5}$ \citep{Cooke2014}. This value of [D/H] is in good agreement with BBN model predictions based on cosmic microwave background observations of $2.58 \pm 0.13 \times 10^{-5}$ \citep{Cyburt2016}.

% its use as a trace of galactic chemical evolution.
If [D/H] can be accurately measured in different locations through the Galactic disk, it should be an extremely valuable yet relatively simple probe of Galactic chemical evolution. Models of the formation of the Galactic disk generally predict that less star formation has occurred in the outer galaxy, and thus that [D/H] should increase with Galactocentric radius (\rgc) \citep[e.g.][]{Chiappini2002,Romano2003,Romano2006,vandeVoort2018}. This picture is complicated by other effects, such as the infall of pristine and deuterium-rich gas which has experienced little astration \citep{Prantzos1996,Romano2006,vandeVoort2018}, and infall from metal-rich/deuterium-depleted gas originating in winds driven by starbursts \citep{Tsujimoto2010,Tsujimoto2011}. To date, however, there have been few direct constraints on [D/H] outside a few kpc from the Sun with which to constrain [D/H] throughout the Galaxy. 

% [D/H] for GCE and the UV measurement and depletion problems. 
Early attempts at measuring [D/H] as a function of \rgc~were made using radio-wavelength transitions of deuterated molecules such as DCN/HCN \citep{Penzias1977,Penzias1979}, however, such surveys are now known to suffer from chemical fractionation effects which can alter the abundance of deuterated molecules by several orders of magnitude \citep{Wootten1987,Awad2014,Ceccarelli2014,Tielens2013}. To date, most measurements of [D/H] in our Galaxy come from UV absorption lines of D I and H I, however, their interpretation in terms of astration is debated (see \citealt{Tosi2010} for a review). Along sight lines with low total hydrogen column density within the local bubble ($\log (N_\mathrm{H}) < 19.2$ cm$^{-2}$), [D/H] is approximately constant at $1.5 \pm 0.4 \times 10^{-5}$, however, beyond the local bubble, variations of a factor of 4-5 in the derived [D/H] are seen between sight lines separated by only a few hundred pc \citep{Linsky2006}. \cite{Linsky2006} have argued that this variation is in fact the result of variations in the depletion of the gas-phase deuterium onto dust grains. Since the C-D bond is stronger than the D-H bond, it is energetically favorable for deuterium to replace hydrogen in carbonaceous dust grains or polcyclic aromatic hydrocarbons \citep[PAHs,][]{Jura1982,Tielens1983,Draine2006,Tielens2008}. Indeed, the C-D bond of aromatic and aliphatic carbon molecules has been detected in emission \citep{Peeters2004, Peeters2024}. \cite{Linsky2006} show that [D/H] along the different sightlines is anti-correlated with the depletion of different refractory elements (e.g. Fe, Si). Regions with a high [D/H] and low levels of depletion may recently have experienced erosion or destruction of dust grains from supernova shocks, thus releasing D back into the gas phase. A tentative correlation between [D/H] and the H$_2$ rotational temperature is also observed, which \cite{Linsky2006} suggest may be the result of clouds with warmer gas having a lower level of refractory depletion. 

If depletion onto dust grains is significant, the highest [D/H] values observed are most representative of the total deuterium abundance in the local galactic disk. Taking dust depletion into account, a Bayesian analysis of the UV measurements suggest the corrected true [D/H] in the local Galactic disk is $ \ge 2.0 \pm 0.1 \times 10^{-5}$ \citep{Prodanovic2010}. This result is supported by recent updated measurements of [D/H] which have been improved with more precise H I measurements, although the correlation between refractory element depletion and [D/H] is somewhat weaker \citep{Friedman2023}. The depletion scenario has also been argued against by \cite{Hebrard2010}, who instead derive [D/H]  from measuring [D/O] with UV absorption lines of D I and O I, and then scaling by the ISM oxygen abundance. They find that [D/O] is in fact homogenous along a similar spread of sightlines, and suggest [D/H] of the local galactic disk may be a much lower value of $7 \pm 2 \times 10^{-6}$ instead. Regardless, UV absorption lines become impractical with distances from the Sun $>2.5$ kpc due to blending of the D I/H I lines that occurs with a higher Hydrogen column density \citep{Hoopes2003}, increasing extinction in the Galactic plane, and the lack of bright background stars at large distances, and are thus also not well-suited to surveys over a wide range of \rgc. 

An alternative simple probe of [D/H] is provided by mid- to far-infrared rotational lines of H$_2$ and HD, the simplest deuterated molecule. In the dense environments of molecular clouds, most deuterium should be in the form of HD, and the measurement of HD lines should provide direct and accurate means of obtaining the [D/H] ratio. As opposed to more complex deuterated species, chemical fractionation effects are not expected to be significant, though the chemical conversion of HD to atomic D can be important for abundance determination \citep{Bertoldi1999}. However, the HD molecule is difficult to observe in the mid-IR due to the low Einstein coefficients $A_\mathrm{ul}$  of order $10^{-5}$ s$^{-1}$ relative to other IR-active species such as CO$_2$ ($A_\mathrm{ul}$ $\sim$ 1-100 s$^{-1}$), high upper energy levels ($E_\mathrm{up} \ge 1900$ K) requiring strong excitation conditions, and its low abundance relative to H$_2$. Furthermore, rotational lines of HD are found in the mid to far-IR, and are not accessible from the ground. Only a handful of such HD detections have thus ever been made with the Infrared Space Observatory (ISO), { \it Spitzer}, and { \it Herschel}. All of these detections are in relatively nearby regions, either in shocks found in protostellar outflows \citep[$< 500$ pc][] {Bertoldi1999,Neufeld2006b,Giannini2011,Yuan2012}, photo dissociation regions \citep{Wright1999}, supernova remnants \citep[$<2$ kpc][]{Yuan2012}, or protoplanetary disks \citep[$< 160$ pc][]{Bergin2013,McClure2016}. A very tentative detection of a single HD transition with ISO made towards Sgr B2 in the Galactic center was claimed by \cite{Polehampton2002}, which will not be discussed further here. 

The James Webb Space Telescope (JWST) now provides a unique opportunity to survey the [D/H] abundance throughout the Galaxy using MIRI/MRS to observe rotational lines of HD and H$_2$ with a single instrument. The sensitivity and spatial resolution of MIRI/MRS (0.3-1.1\arcsec) is greatly improved compared to previous mid-infrared facilities, allowing the H$_2$ and HD column densities and excitation conditions to be spatially mapped in great detail. Moreover, the MIRI/MRS range (4.9-28.5 $\mu$m) covers a variety of shock-tracing lines from iron and sulfur that are accessible in the mid-IR, allowing a comparison of [D/H] and the HD column densities across a variety of different shock environments and providing a potential means to test the dust depletion hypothesis. Finally, the greatly improved sensitivity of JWST allows the faint HD lines to be detected out to much larger distances in the Galaxy than possible before, potentially allowing the [D/H] abundance across the Galaxy to be better constrained.

In this paper, we report the detections of HD lines in 5 nearby bright protostellar outflows and 5 distant high mass protostars at Galactocentric radii (\rgc) from 4 to 11 kpc in the JWST Observations of Young protoStars (JOYS) program (Sect. \ref{sec:obs}). We compare the emission morphology and measure fluxes for H$_2$, HD, and various lines tracing refractory elements in high-velocity jets ([Fe II], [S I]), use rotational diagrams to measure the H$_2$ and HD column densities or upper limits, and determine [D/H] abundance towards our sources in Sect. \ref{sec:analysis}. In Sect. \ref{sec:results}, we compare the observed line fluxes and properties of the H$_2$ and HD gas, and compare the [D/H] abundance between our sources and with Galactic chemical evolution models. We discuss the effects of the depletion of deuterium onto the dust grains, the possible variations in [D/H] across the Galaxy in the context of the Galactic chemical evolution models, and the outlook for future JWST studies of HD in Sect. \ref{sec:disc}. A summary and conclusions are provided in Sect. \ref{sec:conc}

\section{Observations}
\label{sec:obs}

Our MIRI/MRS data are from the JOYS programs 1290 (PI: E. F. van Dishoeck) and 1257 (PI: T. Ray), which target embedded protostars with the pointings generally centered on the continuum position or blue-shifted outflow lobes. Here, we included observations of 5 high-mass star-forming regions, and 5 nearby and bright low-mass sources where the outflows are well resolved. The details of the observations for each source are listed in Table \ref{tab:jwst_targets}. We select low-mass targets with multiple pointings which are bright in H$_2$ so that non-detections of HD can provide significant upper limits on [D/H], and all high-mass targets so that a wide range of Galactocentric radii are sampled, including IRAS 23385+6053 at 11 kpc and several targets in the inner Galaxy at $\sim4$ kpc. We exclude the high-mass source G28-S, as no continuum source is detected even up to the longest MRS wavelength (28 $\mu$m). 

\begin{table*}[htb]
\centering
\caption{Sources analyzed for JWST MIRI/MRS H$_2$ and HD Observations}
\label{tab:jwst_targets}
\begin{tabular}{lcccccc}
\toprule\toprule
Source        &    $d$ (pc) &   $R_\mathrm{GC}$ (pc) & RA (J2000)   & Dec (J2000)   & Integration time (s)\tablefootmark{1} & Pointings \\
\midrule
High-mass          & & & & & \\
\midrule
IRAS 18089-1732     & 3600 & 4600 & 18:11:51.46 &  -17:31:28.8  & 600 & 2 \\
G28 IRS 2           & 3700 & 4700 & 18:42:51.98 &  -03:59:54.0  & 600 & 1\\
G28 P1              & 3700 & 4700 & 18:42:50.70 & -04:03:14.1   & 600 & 1\\

G31-A               & 3750 & 5200 & 18:47:34.32 &  -01:12:46.0  & 600 & 1\\
IRAS 23385+6053 (N) & 4900 & 11000 & 23:40:54.48  & +61:10:27.9  & 600 & 4\\
\midrule
\midrule            
Low-mass          & & & & & \\
\midrule             
L1448-mm         & 293 & $\sim8.0$ & 03:25:38.76  & +30:44:8.2 & 600 & 3\\
IRAS4B           & 293 & $\sim8.0$ & 03:29:12.01 & +31:13:7.1 & 600 & 2\\
BHR71-IRS1       & 199 & $\sim8.0$ & 12:01:36.55 & -65:08:53.6 & 600 & 9\\
Ser-emb-8N       & 436 & $\sim8.0$ & 18:29:49.27 & +01:16:52.6 & 600 & 1\\
HH 211           & 321 & $\sim8.0$ & 03:43:58.10 & +32:00:43.0 & 600 & 24\\
\midrule
\end{tabular}
\tablefoot{The distance relative to the solar system and the Galactocentric radius are denoted $d$ and $R_\mathrm{GC}$ respectively.} The positions given are the pointing centers of the MIRI/MRS observations, which are generally offset from the protostar positions and centered on the blue outflow lobes \tablefoottext{1}{Total integration time of all three gratings per pointing.}
%\tablebib{}
\end{table*}

The data reduction details for HH 211 are described in \cite{CarattioGaratti2024}. For all other sources in program 1290, the reduction details are as follows. The data were taken using the FASTR1 readout mode and a two-point dither pattern in the negative direction which was optimized for extended sources. All three gratings (A, B, C) were used, covering the full 4.9--28.5 $\mu$m wavelength range of MIRI-MRS. The number of pointings varies depending on the angular extent of the source and is listed in Table \ref{tab:jwst_targets}. Dedicated background observations were taken for all sources except G28 IRS 2 and G28 P1 to allow for a subtraction of the telescope background and detector artifacts. However, for IRAS 23385+6053, the background was not subtracted using the dedicated background since it also contains real astronomical continuum emission and PAH features \cite[for details, see][]{Beuther2023}. Since the focus of this paper is on emission lines, the subtraction of telescope background does not significantly alter the results. An astrometric correction was performed for IRAS 23385+6053 using GAIA DR3 stars in the parallel imaged field \citep{Beuther2023}.

The data were processed through all three stages of the JWST calibration pipeline \citep{Bushouse2024} using reference context {\tt jwst$\_$1210.pmap} of the JWST Calibration Reference Data System \citep[CRDS;][]{Greenfield2016}. Both the science and background data were first processed through the {\tt Detector1Pipeline} using the default parameters. Next, the data were further processed through {\tt Spec2Pipeline}, which included the subtraction of the dedicated backgrounds for G31 and IRAS 18089-1732 as well as applying the fringe flat for extended sources (Crouzet et al. in prep.) and the residual fringe correction on the detector level (Kavanagh et al. in prep.). A bad-pixel routine was applied to the resulting {\tt cal} files using the Vortex Imaging Processing (VIP) package \citep{Christiaens2023}. Lastly, the final data cubes were constructed for each subband of each channel individually using the {\tt Spec3Pipeline}. In this final step, the master background and outlier rejection steps were switched off.

\section{Analysis}
\label{sec:analysis}

\subsection{Integrated line intensity maps}
\label{sec:line_maps}

Spatially extended emission from many H$_2$, HD, and atomic lines is detected towards all of our sources (Table \ref{tab:transitions} in App. \ref{sec:app_transitions}). The detected HD lines include transitions with much higher excitation energies than probed by previous mid- and far-IR observations, and range from R(4) ($E_\mathrm{up} =  1895$ K) as high as the R(9) transition ($E_\mathrm{up} = 6656$ K). To compare the morphology of the line emission, it is useful to inspect maps of the continuum subtracted and wavelength integrated line intensity. However, the MIRI/MRS point-spread-function (PSF) size varies significantly across the full wavelength range (0.3-1.1\arcsec), complicating the comparison. Furthermore, individual spaxels in the reconstructed MIRI/MRS data cubes can show amplitude rippling along the spectral axis from undersampling of the PSF at short wavelengths and residual fringing not removed by the pipeline. To enable an even comparison of the line intensities and mitigate these artefacts, we perform a Gaussian convolution of each data cube to a resolution of 1\arcsec, which is approximately the spatial resolution of MIRI/MRS at the wavelength of the longest analyzed line, [S I] at 25.25 $\mu$m. 

We estimate the continuum emission underlying each emission line by taking the median intensity over the wavelength axis in a 1000 km s$^{-1}$ wide range centered around the line's rest wavelength. For each line, we then subtract the estimated continuum and integrate over the wavelength axis in a 400 km s$^{-1}$ wide range centered on the line. This integration range is chosen to capture the full extent of the line because the velocity resolution of MIRI/MRS is only $\sim$100-200 km s$^{-1}$.

\begin{figure*}
    \centering
    \includegraphics[width=0.9\textwidth]{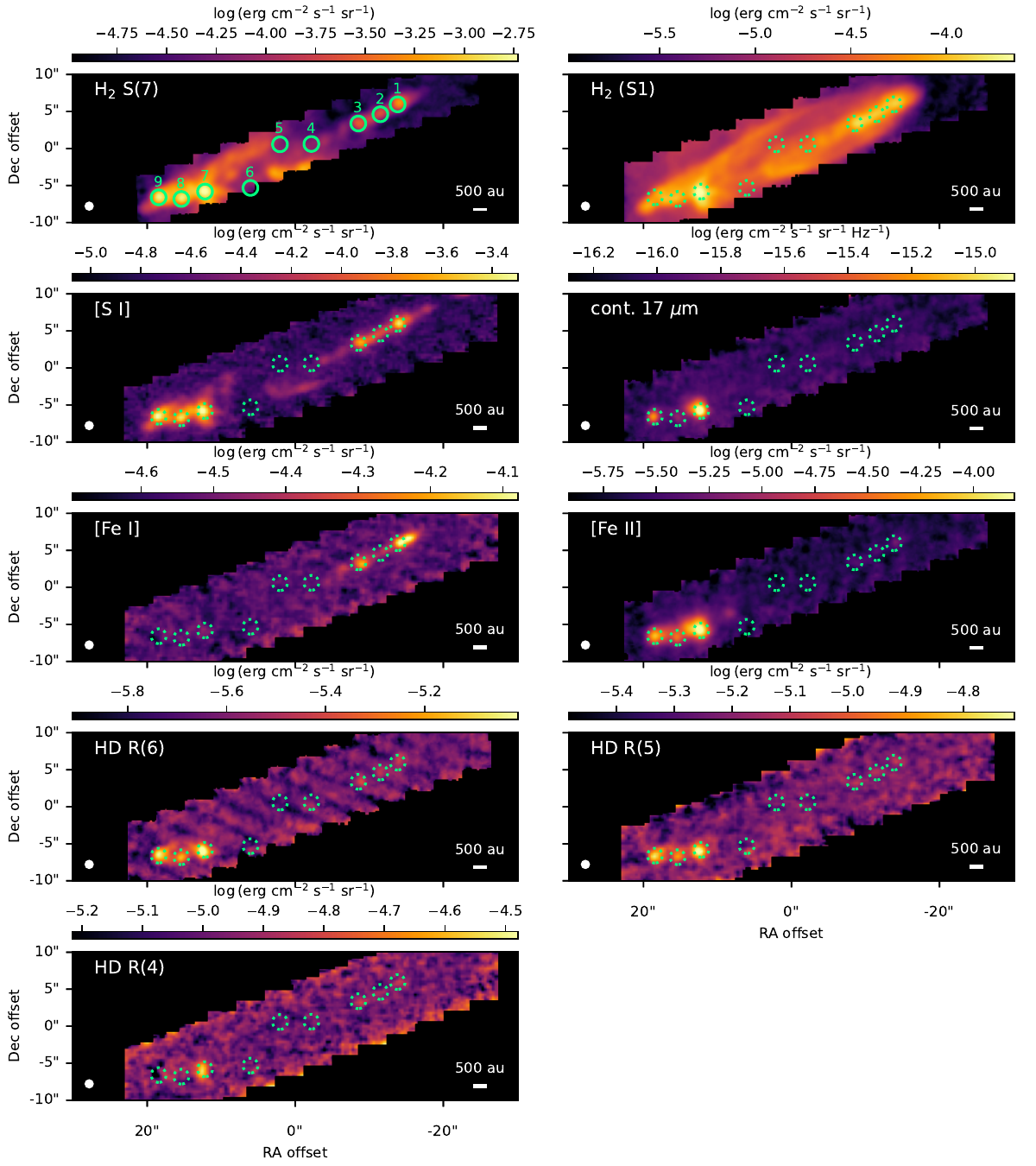}
    \caption{Integrated line intensity maps for HH 211  of various lines and the continuum at 17 $\mu$m shown with a logarithmic stretch. The maps have been smoothed to a common resolution of 1\arcsec, shown by the white circle in the bottom-left. Apertures used for spectral extraction are shown by the green circles. An index for each aperture is provided in the top-left panel. The coordinates of the aperture centers can be found in Table \ref{tab:apertures} of App. \ref{sec:app_apertures}.}
    \label{fig:moment0_1}
\end{figure*}

\begin{figure*}
    \centering
    \includegraphics[width=0.9\textwidth]{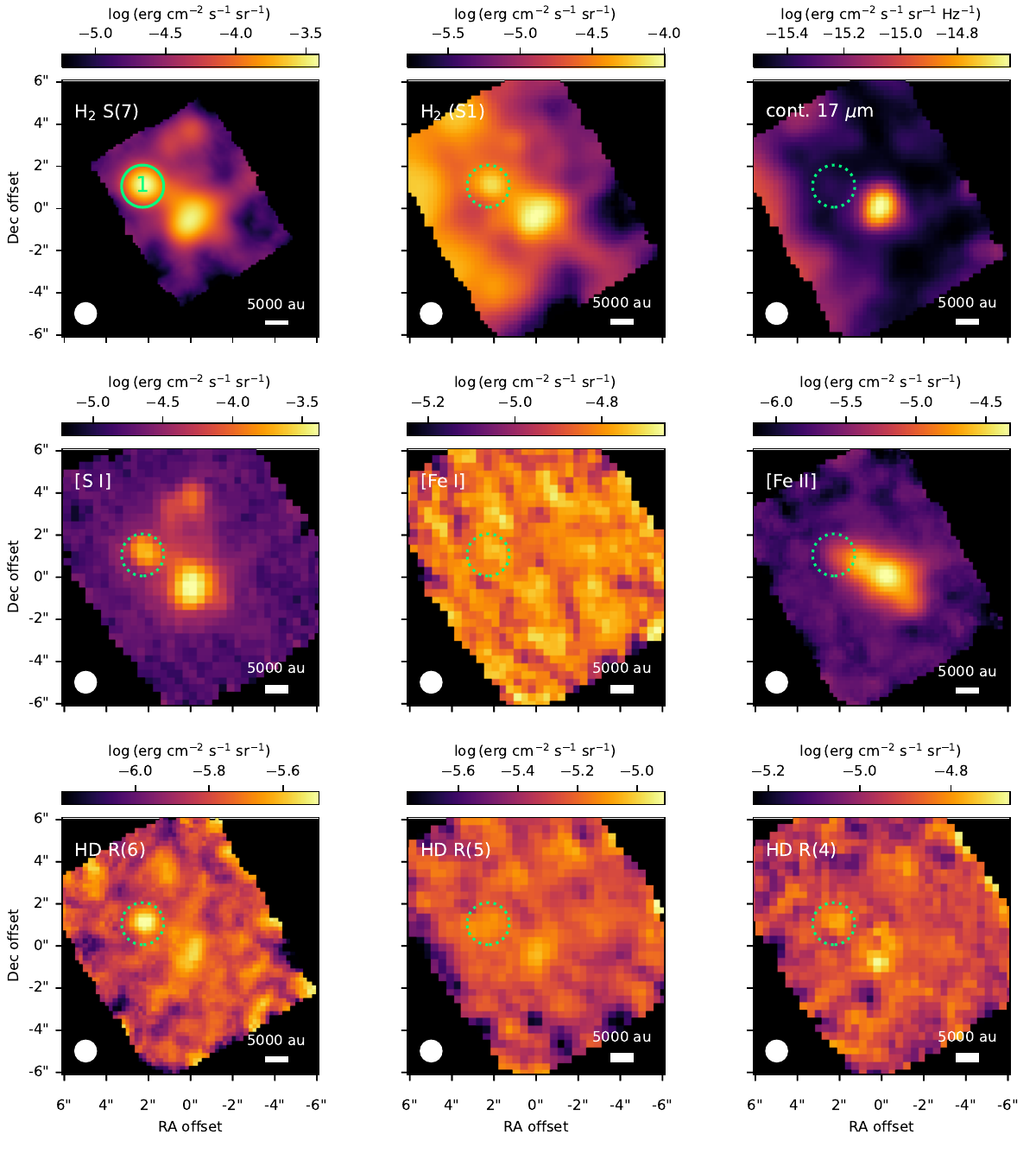}
    \caption{Same as Fig. \ref{fig:moment0_1}, but for IRAS 23385+6053.}
    \label{fig:moment0_2}
\end{figure*}

We show integrated line intensity maps of selected H$_2$, HD, and atomic lines for IRAS 23385+6053 and HH 211 for selected lines in Figs.  \ref{fig:moment0_1} and \ref{fig:moment0_2}, the remaining maps and sources are shown in Figs.  \ref{fig:moment0_3} to \ref{fig:moment0_10}. Where present, emission from HD is seen to be brightest towards bright shocks where the high-J H$_2$ lines are detected.

In HH211, the locations where the HD emission peaks are strongly correlated with the [S I] and [Fe II] atomic lines tracing the high velocity inner jet and bow shock. Emission from [Fe I] is also bright in the inner jet where HD emission is detected, but not in the bow shock, likely reflecting differences in the ionization fraction of Fe. Similar correlations between the HD and atomic line emission are seen for the other nearby low-mass outflows, and for the high mass protostellar system IRAS 23385+6053 where multiple large scale outflows are being driven by embedded protostars, though over far larger scales of 1000s rather than 100s of au \citep{Beuther2023,Gieser2023}. HD emission is not clearly detected in the integrated line intensity maps in any other high-mass sources of the JOYS sample despite the similar sensitivities and distances.

\subsection{Aperture spectral extraction}
\label{sec:line_fluxes}

To quantitatively compare the line fluxes between different sources and outflow regions, we extract spectra from the convolved cubes described in Sect. \ref{sec:line_maps}. From inspection of the integrated line intensity maps in the low-mass sources, the HD lines appear brightest in the outflow jet knots and bow shock, but there are also many outflow locations with bright H$_2$ emission in the lower energy S(1) to S(3) lines ($E_\mathrm{up}\sim$1000-2500 K) where the HD emission is faint despite a large column density of H$_2$ likely being present. To probe the origin of the HD, we thus extract spectra towards positions representative of the HD-bright jets and bow-shocks, as well as the fainter positions in the outflow cavities. For the nearby low-mass sources, we extract multiple apertures to compare conditions across the outflows, while we use only a single aperture for each high-mass source owing to their much greater distance. The locations and sizes of our extracted apertures are shown as green circles overlaid on the line intensity maps described in Sect. \ref{sec:line_maps} and App. \ref{sec:app_line_maps}, and we list their positions in Table \ref{tab:apertures} of App. \ref{sec:app_apertures}.

For each position, we extract a spectrum by taking the sum of the flux in a 1\arcsec~radius aperture. As all data cubes have been convolved to the same resolution of 1\arcsec, the aperture radius is kept constant with wavelength. We then apply a residual fringe correction to the 1D spectrum (Kavanagh et. al. in prep). To measure the flux of the H$_2$, HD, and selected atomic lines, we fit each line with Gaussian profile and a first-order polynomial to represent the continuum. When a line is not detected, we take as an upper limit on the flux $F_\mathrm{up} = 3 \sigma_\mathrm{RMS} \mathrm{FWHM}$, where $\sigma_\mathrm{RMS}$ is measured in a small line-free region adjacent to the expected line, and FWHM is the full width at half-maximum of MIRI/MRS spectral response. This empirical approach in determining the upper limits incorporates both the sensitivity limits of MIRI/MRS and calibration uncertainties, e.g. residual high-frequency fringing remaining in the spectrum.

\begin{figure*}[h]
    \centering
    \includegraphics[width=\textwidth]{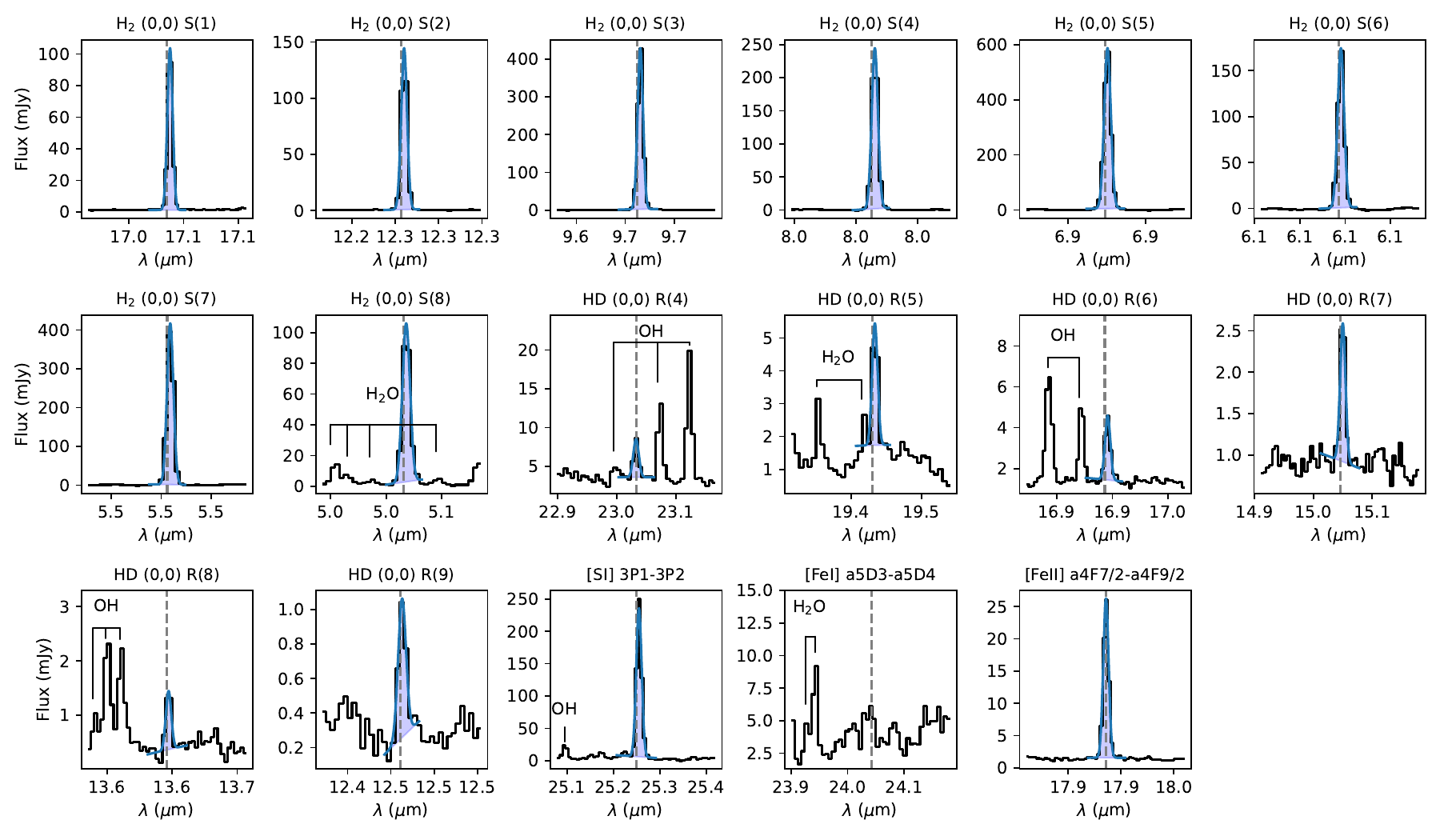}
    \caption{Spectral line fits for H$_2$ (top two rows), HD (rows 3 and 4) , and the [S I], [Fe I], and [Fe II] atomic lines (rows 4 and 5) for HH 211 aperture 8. The line flux (shaded blue region) is determined by the simultaneous fit of a Gaussian profile and a first order polynomial to represent the continuum. the rest-frame wavelength of each spectral line is shown by a dashed grey line, and non-detections are shown without a shaded region for the fit. Nearby molecular lines of are labelled.}
    \label{fig:HH211_line_det}
\end{figure*}

We show line detections and Gaussian fits for two representative apertures from HH 211 and IRAS 23385+6053 in Figs.  \ref{fig:HH211_line_det} and \ref{fig:IRAS23385_line_det}, respectively. The fluxes for all lines and apertures are reported in Tables \ref{tab:h2_line_fluxes} to \ref{tab:atomic_line_fluxes} in App. \ref{sec:app_line_fluxes}. The lines are generally unblended, except for the HD R(6) line which can overlap with a bright transition of H$_2$O in some cases, and the HD R(7) line which can become partially blended with CO$_2$ Q and P-branch transitions. Bright OH transitions are often detected close to the HD R(6) line but do not overlap it.

\subsection{Rotation diagram analysis}
\label{ssec:rotation_diagrams}

We use a rotation diagram analysis to measure the column density of H$_2$ and HD, and thus determine the [D/H] abundance or upper limits for our targets. Under the assumption that the line emission is optically thin and in local thermodynamic equilibrium (LTE), a rotation diagram provides a simple way to determine the column density and kinetic temperature of the gas \citep{Mangum2015}. The extinction-corrected line intensities $I_{ul}$ (explained below) are used to compute the quantity for the y-axis of the diagram
$$y=\ln(N_u/g_u) = \ln\left(\frac{4 \pi I_{ul}}{A_{ul}h\nu g_u}\right),$$
where $N_u$ is the column density of molecules in the upper state of the transition, g$_u$ the upper state degeneracy, $A_{ul}$ the Einstein $A$ emission coefficient, and $h\nu$ the photon energy. The x-axis of the diagram is the upper state energy $x=E_u/k$ in units of K, and on such a diagram a linear fit of the form $y=mx + b$ will yield the column density of the gas and temperature via:
$m = -1/T$ and $b = \ln(N/Q(T)),$ where $N$ is the total column density of the molecule and $Q(T)$ is the partition function.

To convert the wavelength-integrated line fluxes from Sect. \ref{sec:line_fluxes} to the intensity needed for the rotation diagrams, we assume that the H$_2$ or HD emission fills the aperture. This assumption should be reasonable for the nearby low-mass sources, as modeling of the H$_2$ excitation in shocks indicates that most should be spatially resolvable by JWST/MIRI for shock densities $\lesssim 10^6$ cm$^{-3}$ \citep[][Figure 9]{Kristensen2023}. For the distant high-mass sources the emission solid angle may be much smaller than the aperture, but we nonetheless make the same assumption. In any case the assumed emission solid angle is the same for both H$_2$ and HD and thus there is no impact on any column density ratios or the derived [D/H]. The observed line intensity $I_{ul,\mathrm{obs}}$ is then
$I_{ul, \mathrm{obs}} = {F_{ul, \mathrm{obs}}}{\Omega},$
where $F_{ul, \mathrm{obs}}$ is the sum of the integrated line flux in the extraction aperture with solid angle $\Omega $ equal to a circle with radius 1\arcsec. 

The extinction correction needed for the line intensities can be derived using the H$_2$ rotation diagrams and a wavelength dependent extinction law $A_\lambda(A_K)$. In Fig. \ref{fig:ext_curve_and_lines} of App. \ref{sec:ext_corr_app}, we show the KP5 \citep{Pontoppidan2024} extinction curve at $A_K$ values ranging from 1 to 10.
This theoretical extinction curve constrained by observations is appropriate for dense molecular clouds, and contains absorption features from silicates and various ice species. It has been shown to perform well for extinction correction of H$_2$ lines in JWST spectra \citep{Narang2024}.
The $J=1-4$ transitions are particularly useful as a measure of extinction, as the S(3) transition at $\sim9.66$ $\mu$m is found within the broad 10 $\mu$m silicate feature, whereas the S(1), S(2), and S(4) transitions lie outside this feature and experience approximately the same extinction. Using the uncorrected intensities to produce a rotation diagram, the S(3) transition will thus appear relatively suppressed, as has been found in many previous H$_2$ studies \citep[e.g.][]{Bertoldi1999,Neufeld1998}. By fitting a single temperature component with a variable $A_K$ to the S(1) to S(4) transitions, the rotation diagram is effectively linearized. Specifically, a fit to the rotation diagram using the uncorrected intensities is done with the relation:
$$y_\mathrm{obs} = \ln\left( N_\mathrm{warm}  \frac{\exp(-x / T_\mathrm{warm})}{Q(T_\mathrm{warm})} 10^{-A_\lambda/2.5}  \right),$$
where
$$y_\mathrm{obs} = \ln\left(\frac{4 \pi I_{ul,\mathrm{obs}}}{A_{ul}h\nu g_u}\right).$$

We perform this fit with equal weighting on all lines because of the importance of the relatively faint S(3) line in the determination of the extinction. We show the effects of this correction for HH 211  and IRAS 23385+6053 in Fig. \ref{fig:ext_corr_example} in App. \ref{sec:ext_corr_app}.

After correction for extinction, our H$_2$ rotation diagrams all show some amount of curvature in the higher upper energy S(5)-S(8) lines and are thus not well described by a single temperature component. Optically thick emission can produce curvature in a rotation diagram, but this should not be significant for H$_2$ or HD rotational emission in protostellar outflows given the low column density of the gas and small Einstein coefficients. A more reasonable explanation is the presence of multiple temperature components in different shock environments in the aperture. This also produces curvature in a rotation diagram, and has been noted in many previous studies of H$_2$ emission from outflows \citep{vanDishoeck2004,Yuan2011,Yuan2012}.
 In addition to curvature, some of our H$_2$ rotation diagrams show a zigzag pattern where the ortho (odd-J) transitions are suppressed relative to the para (even-J) transitions (See Fig. \ref{fig:opr_corr_example}). This is characteristic of an ortho-to-para H$_2$ column density ratio less than the typical value of 3 reached in local thermodynamic equilibrium \citep{Neufeld2006a}.

\begin{figure*}[h]
    \centering
    \includegraphics[width=0.24\textwidth]{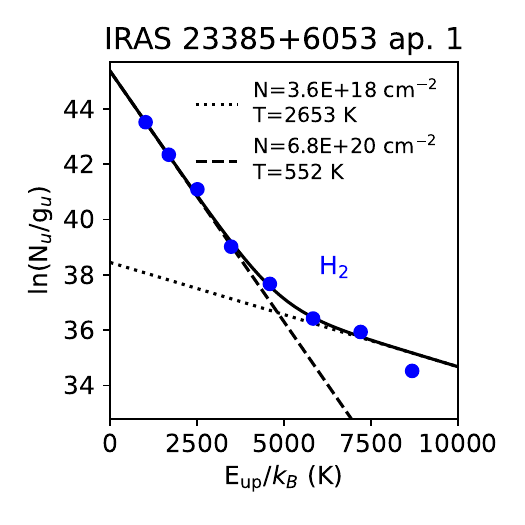}
    \includegraphics[width=0.24\textwidth]{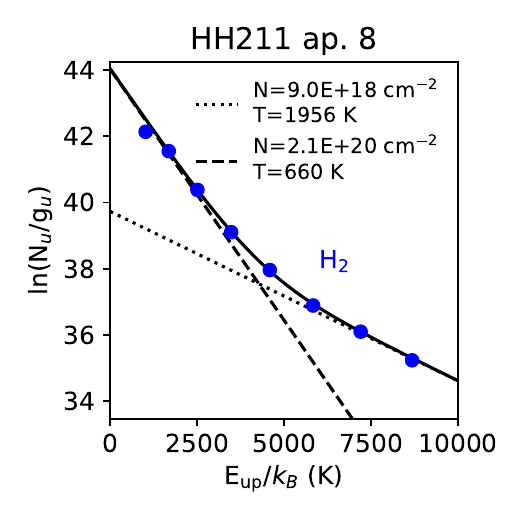}
    \includegraphics[width=0.24\textwidth]{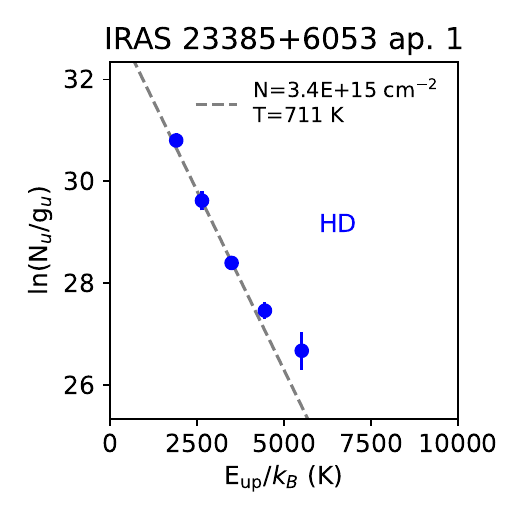}
    \includegraphics[width=0.24\textwidth]{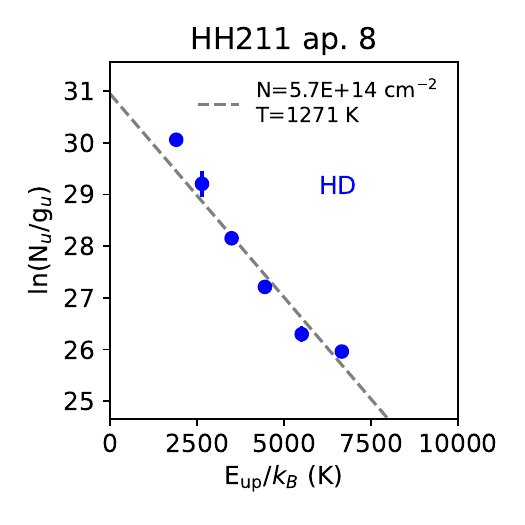}
    \caption{Rotational diagrams for the detected H$_2$ (top row) and HD (bottom row) transitions in IRAS 23385+6053 aperture 1 and HH 211  aperture 8. Corrections for extinction and an ortho-to-para ratio $<3$ have been applied (see text and App. \ref{sec:ext_corr_app}).}
    \label{fig:main_rot_diagrams}
\end{figure*}

To fit the extinction corrected H$_2$ rotation diagrams, we thus use a two-temperature model with an additional correction for the ortho-to-para correction term: 
$$y= \ln\left( N_\mathrm{warm}  \frac{\exp(-x / T_\mathrm{warm})}{Q(T_\mathrm{warm})} + N_\mathrm{hot}  \frac{\exp(-x / T_\mathrm{hot})}{Q(T_\mathrm{hot})}  \right) + z(J),$$
where $N_\mathrm{tot} = N_\mathrm{warm} + N_\mathrm{hot}$, for a warm component ($T_\mathrm{warm}=250-1000$ K) and hot component ($T_\mathrm{hot}=1000-6000$ K), and where 
$$
    z(J)= 
\begin{cases}

    \ln\left(\frac{\mathrm{OPR}}{3}\right)& \text{odd } J\\
    0,              & \text{even } J
\end{cases},
$$
and $\mathrm{OPR}$ is the ortho-to-para ratio. Representative fits for IRAS 23385+6053 and HH 211 are shown in the top row of Fig. \ref{fig:main_rot_diagrams}. We summarize the H$_2$ fit results in Table \ref{tab:rot_diagram_results}. We note that the error bars on the fit parameters include only the uncertainty propagated from the line flux measurements; we consider the effect of uncertainty in $A_K$ on deriving [D/H] in Sect. \ref{ssec:DtoH_measurement}.

The majority of our H$_2$ rotation diagrams are well described by two temperature components in LTE with $\mathrm{OPR}=3$. In some cases, $\mathrm{OPR}=2-3$ is found, but the values of the warm and hot components are still consistent with LTE. For a handful of positions associated with very bright bow-shocks, the rotation diagrams become nearly horizontal for the higher excitation transitions (S(4) to S(8)). Here, the temperature of the hot component is very poorly constrained and reaches the maximum allowed in the fit of 3000 K. This suggests a still hotter component of the H$_2$ emission, which has been detected in rovibrational emission from other outflows with ISO-SWS and JWST/NIRspec \citep{vanDishoeck1998,Giannini2008,Giannini2015,LeGouellec2024}.
In any case, the column density of the warm H$_2$ component which we use in our estimate of [D/H] is determined almost entirely by the S(1) to S(4) lines, for which the temperature is well constrained.

The rotation diagram fits for HD are somewhat simpler, as we see little evidence of multiple temperature components, and HD does not have ortho and para forms. Since the HD is also only present in the outflow, it should experience the same extinction as H$_2$, and we thus correct the line intensities using the same value for $A_K$, and perform a fit of a single temperature component to the rotation diagram.Representative fits for IRAS 23385+6053 and HH 211 are shown in the bottom row of Fig. \ref{fig:main_rot_diagrams}. We note that the excitation energies of the HD R(4) to R(6) lines are similar to the H$_2$ S(2) to S(4), so the single component fit to the HD transitions should trace gas with similar excitation conditions as the warm component in the H$_2$ fits, which dominates $N_\mathrm{tot}$. However, the much larger Einstein coefficients of the HD transitions than those of H$_2$ result in a higher critical density, and deviations from LTE may become significant for the low densities in protostellar outflows. We discuss this further in Sect. \ref{ssec:DtoH_measurement}, where we correct for non-LTE conditions in our determination of [D/H]. We note that we do not consider a temperature gradient as in previous works analyzing Spitzer detections of HD \citep{Yuan2011,Yuan2012}, as the HD lines analyzed here are not sensitive to the cold ($< 300$ K) gas that could be probed with the lower excitation HD lines detectable with Spitzer.

For our other JOYS targets where H$_2$ is detected but no HD lines are observed, we can estimate upper limits on the HD column density. We convert each HD line upper limit to a column density assuming the temperature of the warm component of H$_2$ in the rotation diagram fits, and applying the same extinction correction to the line intensities as was derived for H$_2$. We take the lowest column density upper limit of the three transitions as our overall 3$\sigma$ upper limit for each observation. We note that despite the much lower sensitivity in channel 4 of MIRI/MRS, the most constraining upper limit is typically from the 23.04 $\mu$m R(4) line of HD, as this transition lies at the lowest upper energy level (see Table \ref{tab:transitions}). We summarize the results of our HD rotation diagram fits in Table \ref{tab:rot_diagram_results}. Where we have detections of HD, 3-5 transitions are detected, which provides much more precise constraints on the column density and temperature of HD in protostellar outflows than possible with previous facilities. In general, we find the excitation temperatures of the single component for HD to be intermediate between the temperatures of the warm and hot H$_2$ components, typically 500-1000 K, and the column density of HD to be $\sim5$ orders of magnitude lower than that of the warm H$_2$ component. We can thus rule out that enhancement of HD by orders of magnitude relative to H$_2$ is occurring, as is often the case for other complex deuterated species \citep{Ceccarelli2014}.

\subsection{[D/H] abundance and upper limits}
\label{ssec:DtoH_measurement}

As a starting point to determine [D/H] abundances, we can assume LTE holds, and compare the column density from the warm component of H$_2$ with that of HD traced the same approximate ranges in upper energy of the transitions. For our protostellar outflows, we assume that the outflows are almost entirely molecular H$_2$, except for a core of atomic H associated with the jet, as has been found for HH 211 \citep{CarattioGaratti2024}. The [D/H] abundance or upper limit is then simply $\mathrm{[D/H]} = \frac{1}{2} N_\mathrm{HD}/N_\mathrm{H_2}$. However, at densities $< 1 \times 10^6 $ cm$^{-3}$ that can occur in protostellar outflows, non-LTE deviations in the level populations for the $J>4$ HD lines are significant \citep{Bertoldi1999}. Prior estimates of the H$_2$ from a non-LTE analysis of a handful of detected HD transitions towards the L1448 outflow suggest that densities $< 1 \times 10^6 $ cm$^{-3}$ are indeed present in some regions of the flow \citep{Giannini2011}. \cite{Bertoldi1999} show that not accounting for the non-LTE excitation can result in an overestimate of the HD column density. For their analysis of HD in the Orion OMC-1 outflow, they found that non-LTE excitation increases the inferred column density by a factor of $\sim 1.47$ relative to an LTE model, though the exact correction factor is sensitive to the shock density, temperature, and dissociation fraction.

Another important effect on the derived deuterium abundance considered by \cite{Bertoldi1999} is the chemical conversion of HD to D via: HD + H + $\Delta H_0$ $\rightleftharpoons$ D + H$_2$, where $\Delta H_0 = 418$ K is the enthalpy difference. This can be significant in the warmest and densest parts of C-type shocks present in some regions of protostellar outflows, though HD reforms in the cooler post-shock regions. For the Orion OMC-1 outflow, \cite{Bertoldi1999} found that the average effect over the entire shock region was to reduce the HD abundance relative to H$_2$ by a factor of 1.67. When the effect of chemical conversion is combined with the non-LTE excitation, their true [D/H] abundance is a factor of $2.45$ higher than what would be inferred from a purely LTE analysis. Interestingly, \cite{Bertoldi1999} find that this result is not particularly sensitive to how dissociative the shock is. In a less dissociative shock with a smaller fraction of atomic H, chemical conversion of HD to D is weaker but the correction for non-LTE effects is larger, and the combined correction factor is approximately constant. To account for the combined effects of chemical conversion and non-LTE excitation of HD, we thus follow \cite{Bertoldi1999} and multiply our LTE [D/H] estimates by a factor of 2.45.

An estimate for the uncertainty for $\mathrm{[D/H]}$ should also include the effect of extinction. However, the uncertainty in the column densities of warm H$_2$ and HD from Sect. \ref{ssec:rotation_diagrams} includes only the propagated errors on the line fluxes, as $A_K$ was held fixed to the value determined from the S(1) to S(4) transitions of H$_2$. To estimate the effect of $A_K$ uncertainty on the derived $\mathrm{[D/H]}$, we rerun our two component fits with $A_K$ up to two magnitudes larger or smaller than the best fit value. As an example, we show this procedure for IRAS 23385+6053 in Fig. \ref{fig:extinction_uncertainty} in App. \ref{sec:ext_corr_app}. The resulting column densities for warm H$_2$ and HD are shown as a function of $A_K$ in the upper panel, while the derived $\mathrm{[D/H]}$ is shown in the lower panel. Since both the HD and H$_2$ column densities are strongly correlated with any change in $A_K$, the overall effect on $\mathrm{[D/H]}$ is small. We take as our uncertainty in $A_K$ the result from the fit of a single temperature component to the S(1) to S(4) transition (green shaded region in lower panel of Fig. \ref{fig:extinction_uncertainty}), and convert this to an uncertainty in $\mathrm{[D/H]}$ (blue shaded region in lower panel of Fig. \ref{fig:extinction_uncertainty}). We include the uncertainty of $A_K$ it in our final estimate of $\mathrm{[D/H]}$ by adding it in quadrature with the uncertainty propagated from the fits to the rotation diagrams.

\section{Results}
\label{sec:results}

We summarize our inferred deuterium abundances or upper limits in Table \ref{tab:DtoH}, and also list the Galactocentric radius \rgc~  to each source. For apertures and sources with a non-detection of HD and low column densities and/or low temperatures of the warm H$_2$ component, the upper limits on the HD column density result are low as the HD lines are not expected to be bright. This results in corresponding upper limits on [D/H] which are well above the primordial abundance of $2.58 \pm 0.13 \times 10^{-5}$ \citep{Cyburt2016}, and thus not constraining. 

\begin{table}[htb]
\caption{[D/H] Abundances from mid-IR rotational lines}
\label{tab:DtoH}
\begin{tabular}{lrll}
\toprule\toprule
                Source + Aperture &  $R_{GC}$ &           [D/H] $\times 10^{-5}$ &                ref. \\
\midrule
G28-IRS2 ap. 1 & 4.7 & $< 4.81$ & JOYS \\
G28-P1-A ap. 1 & 4.7 & $< 1.90$ & JOYS \\
G31-A ap. 1 & 5.2 & $< 3.69$ & JOYS \\
IRAS 18089-1732 ap. 1 & 4.6 & $< 0.71$ & JOYS \\
IRAS 23385+6053 ap. 1 & 11.0 & $0.61 \pm 0.30$ & JOYS \\
BHR71-IRS1 ap. 1 & $\sim8.0$ & $1.90 \pm 1.03$ & JOYS \\
BHR71-IRS1 ap. 2 & $\sim8.0$ & $1.06 \pm 0.46$ & JOYS \\
BHR71-IRS1 ap. 3 & $\sim8.0$ & $1.26 \pm 0.25$ & JOYS \\
BHR71-IRS1 ap. 4 & $\sim8.0$ & $< 2.46$ & JOYS \\
HH 211 ap. 1 & $\sim8.0$ & $1.30 \pm 0.49$ & JOYS \\
HH 211 ap. 2 & $\sim8.0$ & $1.74 \pm 0.14$ & JOYS \\
HH 211 ap. 3 & $\sim8.0$ & $1.78 \pm 0.47$ & JOYS \\
HH 211 ap. 4 & $\sim8.0$ & $< 1.17$ & JOYS \\
HH 211 ap. 5 & $\sim8.0$ & $< 0.86$ & JOYS \\
HH 211 ap. 6 & $\sim8.0$ & $< 2.92$ & JOYS \\
HH 211 ap. 7 & $\sim8.0$ & $0.25 \pm 0.18$ & JOYS \\
HH 211 ap. 8 & $\sim8.0$ & $0.33 \pm 0.15$ & JOYS \\
HH 211 ap. 9 & $\sim8.0$ & $0.44 \pm 0.34$ & JOYS \\
IRAS 4B ap. 1 & $\sim8.0$ & $0.12 \pm 0.07$ & JOYS \\
IRAS 4B ap. 2 & $\sim8.0$ & $0.12 \pm 0.06$ & JOYS \\
L1448-mm ap. 1 & $\sim8.0$ & $1.66 \pm 0.42$ & JOYS \\
L1448-mm ap. 2 & $\sim8.0$ & $2.27 \pm 0.21$ & JOYS \\
L1448-mm ap. 3 & $\sim8.0$ & $2.24 \pm 0.36$ & JOYS \\
L1448-mm ap. 4 & $\sim8.0$ & $< 1.38$ & JOYS \\
Ser-emb 8 (N) ap. 1 & $\sim8.0$ & $0.92 \pm 0.45$ & JOYS \\
Ser-emb 8 (N) ap. 2 & $\sim8.0$ & $0.57 \pm 0.38$ & JOYS \\
Local disk [D I]/[H I] & $\sim8.0$ & $2.00 \pm 0.10$ & $^a$ \\
Local disk [D I]/[O I] & $\sim8.0$ & $0.70 \pm 0.20$ & $^b$ \\
Orion Bar & 8.4 & $1.00 \pm 0.30$ & $^c$ \\
Orion OMC & 8.4 & $0.76 \pm 0.29$ & $^d$ \\
\bottomrule
\end{tabular}
\\All listed JOYS values have been corrected upwards for chemical conversion of HD by a factor of 2.54 (see text). \\
$^a$ \citep{Prodanovic2010}. \\
$^b$ \citep{Hebrard2010}. \\
$^c$ \citep{Wright1999}. \\
$^d$ \citep{Bertoldi1999}.
%\tablebib{}
\end{table}

We also include in Table \ref{tab:DtoH} literature [D/H] measurements that have been derived from rotational lines of HD and H$_2$ observed in a protostellar outflow and photodissociation region (PDR) with ISO \citep{Bertoldi1999,Wright1999}. We note that similar rotation-line-based [D/H] measurements are available from {\it Spitzer} observations of 
two nearby protostellar outflows and a supernova remnant \citep{Yuan2012}, however, 
the \cite{Yuan2012} [D/H] values have systematic uncertainties of a factor of a few depending on whether spectroscopic or imaging data is used to estimate line fluxes, and thus are not particularly constraining. A detection of DCN in a cloud $\sim 0.1$ pc from the Galactic center was reported by \cite{Lubowich2000}, and through comparison with an HCN line and chemical fractionation modelling [D/H]$=1.7\pm0.3\times10^{-6}$ was claimed, however, this result may also vary by an order of magnitude or more given the systematic uncertainty associated with fractionation models and their strong temperature dependence. 
Finally, we also include the [D/H]$=2.0\pm0.1\times10^{-5}$ in the local Galactic disk derived through UV absorption lines of D I and H I \citep{Prodanovic2010}, and the [D/H]=$7.0\pm2.0\times10^{-6}$ derived from D I and O I lines and scaling by the ISM oxygen abundance \citep{Hebrard2010}.

\begin{figure*}[b]
    \centering
    \includegraphics[width=0.8\linewidth]{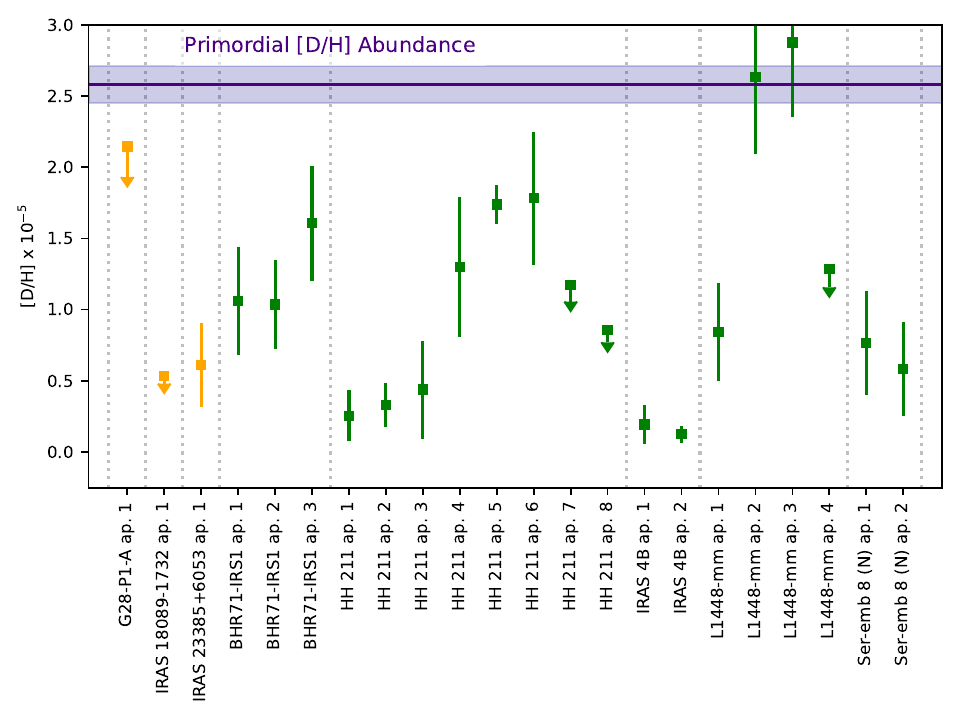}
    \caption{Comparison of [D/H] abundance (Table \ref{tab:DtoH}) in this work. A correction factor of 2.54 has been applied to our [D/H] measurements (see text). Orange symbols show [D/H] in apertures for high-mass sources, while green symbols show [D/H] in apertures from low-mass sources. The dotted grey lines separate different sources. The blue shaded region shows the primordial total [D/H] abundance of \citep{Cyburt2016}. [D/H] upper limits above the primordial abundance are omitted.}
    \label{fig:DtoH_scatter}
\end{figure*}

 Fig. \ref{fig:DtoH_scatter} presents our sources with detected HD lines and a [D/H] measurement (including the upwards correction factor of 2.45 for non-LTE effects and chemical conversion, see Sect. \ref{ssec:DtoH_measurement}), or a [D/H] upper limit below the primordial abundance. Significant variations in [D/H] abundance of a factor $\sim4$ are seen between detected sources. The outflows from low-mass protostars (green symbols in Fig. \ref{fig:DtoH_scatter}) originate from star-forming regions at similar positions in the Galactic disk. These regions should have formed from gas that has experienced similar levels of astration and thus with nearly the same [D/H] abundance (discussed further in Sect. \ref{sec:disc}). This should certainly be true when comparing different positions from the \emph{same} outflow (e.g. HH 211  and L1448-mm), however, a similar magnitude of variation as between different sources is seen instead. We note that in HH 211 [D/H] is robustly lower in the bow-shock positions (apertures 1-3) than the jet positions (apertures 4-6), hinting at a link between the shock environment and [D/H]. Overall, it is thus unlikely that the [D/H] variations seen are the result of differences in the degree of astration for the low-mass sources. However, it is crucial to note that the [D/H] derived from observations, whether from rotational H$_2$ and HD lines or H I and D I absorption lines, is only sensitive to the \emph{gas-phase} component. Galactic chemical evolution (GCE) models make predictions for the \emph{total} [D/H], including both molecular and atomic forms and any deuterium sequestered in dust grains or deuterated PAHs. 

We note that the factor of $\sim 4$ abundance variation is similar to that seen by \cite{Linsky2006} and \cite{Prodanovic2010} when comparing their UV absorption [D/H] measurements along different sight lines in the local Galactic disk, which they attribute to variations in the degree of deuterium depletion onto dust grains. It is thus plausible that the same depletion effect could explain the variations we see between different outflows and outflow positions, which we investigate further in the following Section.

\subsection{Line flux and rotation diagram correlations}
\label{line:corr}

As noted in Sect. \ref{sec:line_maps}, the integrated line intensity maps show that the morphology of the HD emission is clearly associated with the jets and bow-shocks in our sources. The HD emission is particularly bright towards outflow positions that are also bright in [Fe I] and [S I]. Iron and Sulfur are both known to be depleted onto dust grains \citep{Jenkins2009}, though the exact reservoir of the latter is not well-understood \citep{Anderson2013,Kama2019}. A possible explanation for the association between HD, [Fe I], and [S I] is that deuterium is significantly depleted onto dust grains, and is subsequently released along with Fe and S in strong shocks where the dust grains are destroyed. Free atomic D can be readily converted to HD via D + H$_2$ $\rightarrow$ HD + H.

\begin{figure*}[htb]
    \centering
    \includegraphics[width=0.7\textwidth]{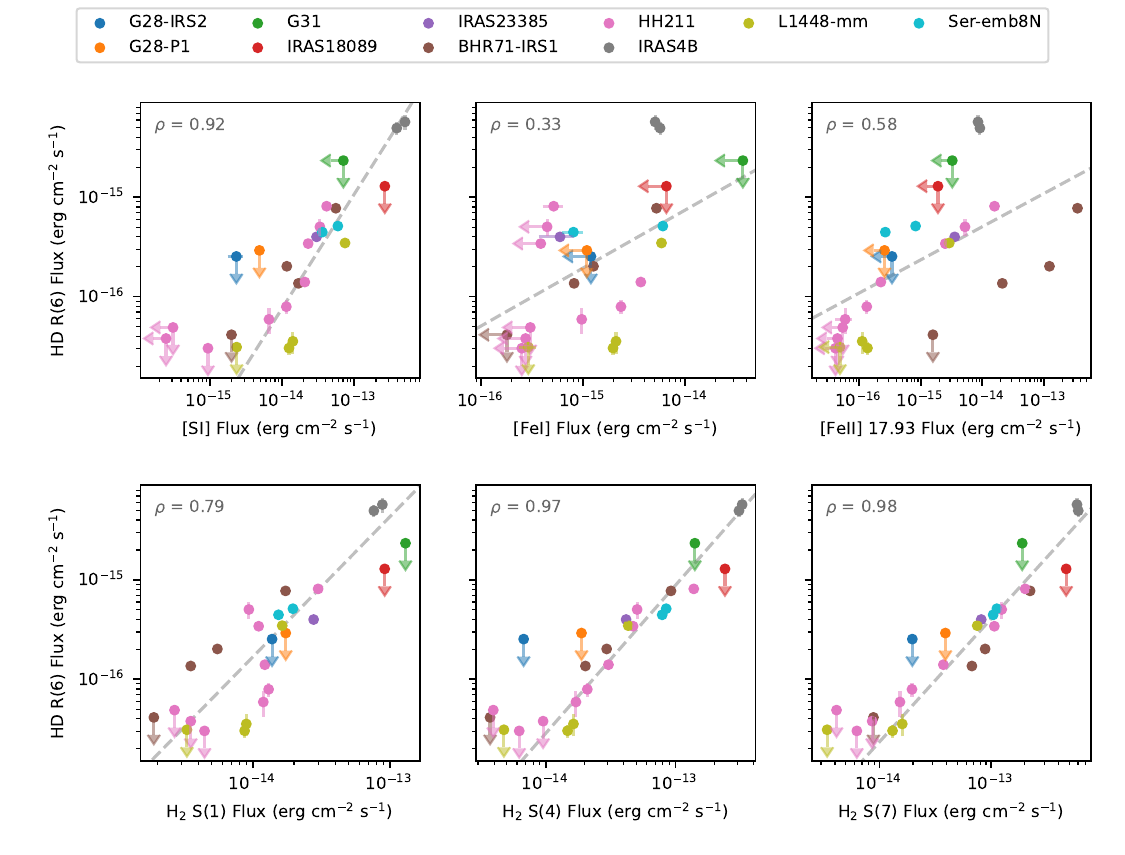}
    \caption{Correlations between the extinction corrected flux of the HD R(6) line measured in all apertures for our sample with the extinction corrected flux of various atomic and H$_2$ lines. The line flux is related to the line intensity by a constant factor of the aperture solid angle (see text). {\it Top row:} Correlation with the [S I], [Fe I], and [Fe II] line flux. {\it Bottom row:} Correlation with the H$_2$ S(1), S(4), and S(7) line flux.}
    \label{fig:line_corr_HD}
\end{figure*}

To perform a more quantitative comparison, we can investigate correlations between the extracted line fluxes and the rotation diagram fit parameters in all of our apertures. We first apply the same extinction correction derived from the H$_2$ rotation diagrams to all of our line fluxes using the method described in Sect. \ref{ssec:rotation_diagrams}. We note that since the emission is extended and should fill the extracted apertures, the line flux and intensity are different only by a constant factor of the aperture solid angle. We show in the top row of Fig. \ref{fig:line_corr_HD} correlations between the HD R(6) line and 3 other lines associated with the outflow jets or bow-shocks: [S I], [Fe I], [Fe II]. A particularly strong correlation ($\rho=0.95$) is seen between the HD R(6) line and [S I], while the correlation with [Fe I] and [Fe II] exhibit much more scatter ($\rho=0.36$ and $\rho=0.57$ respectively). This is possibly due to variations in the ionization conditions between different shocks that affect the intensity of the [Fe I] and [Fe II] lines, but which are not as significant for [S I]. Namely, the ionization potential of the [Fe I] line is only 7.9 eV while for [S I] it is 10.36 eV. In the bottom row of Fig. \ref{fig:line_corr_HD}, we show correlations between the HD R(6) line and the H$_2$ S(1), S(4), and S(7) lines. Strong correlations are seen particularly with the higher upper-energy S(7) line ($E_\mathrm{up}\sim$ 7200 K), suggesting the HD R(6) emission is associated with the denser and higher temperature components of the outflow.

While the association of the HD emission with dense and shocked regions of the outflows is clear, stronger evidence of D depletion and release in shocks would be a correlation between [D/H] and the gas phase abundance of refractory species or some measure of the degree of dust grain destruction. Determining the degree of dust destruction in the outflow shocks is non-trivial, however. For the case of HH211, modeling of the gas-phase abundance of Fe towards bright H$_2$ knots in the inner jet and bow shock has been performed by \cite{CarattioGaratti2024} using the models of \cite{Hollenbach1989}. Their results show that the gas-phase iron abundance in all shock knots are depleted with respect to the solar value, and the knots in the inner jet are less depleted than those in the bow shock. \citep[][Fig. 13]{CarattioGaratti2015}. This is the opposite of the expectation that the gas-phase iron abundance should be higher further along the flow as more grain destruction occurs and more Fe is released \citep{Nisini2005,Podio2006}. However, it is consistent with the increased [D/H] in the inner jet (apertures 4-6) relative to the bow-shocks in HH 211 (apertures 1-3), suggesting depletion of deuterium into dust grains is indeed a plausible mechanism. We note that two important caveats in the modeling of \cite{CarattioGaratti2024} are the uncertainty in the shock velocity and density, which makes the relative Fe depletion somewhat uncertain, and the assumption that sulfur is undepleted in the line ratios used, which affects the absolute gas-phase abundance. To confirm the link between dust depletion and [D/H] variations in the outflow, further modeling of the shock conditions and degree of refractory depletion is needed.

\subsection{[D/H] variations and Galactic chemical evolution}
\label{ssec:DtoH_GCE}

\begin{figure*}[htb]
    \centering
    \includegraphics[width=1.0\linewidth]{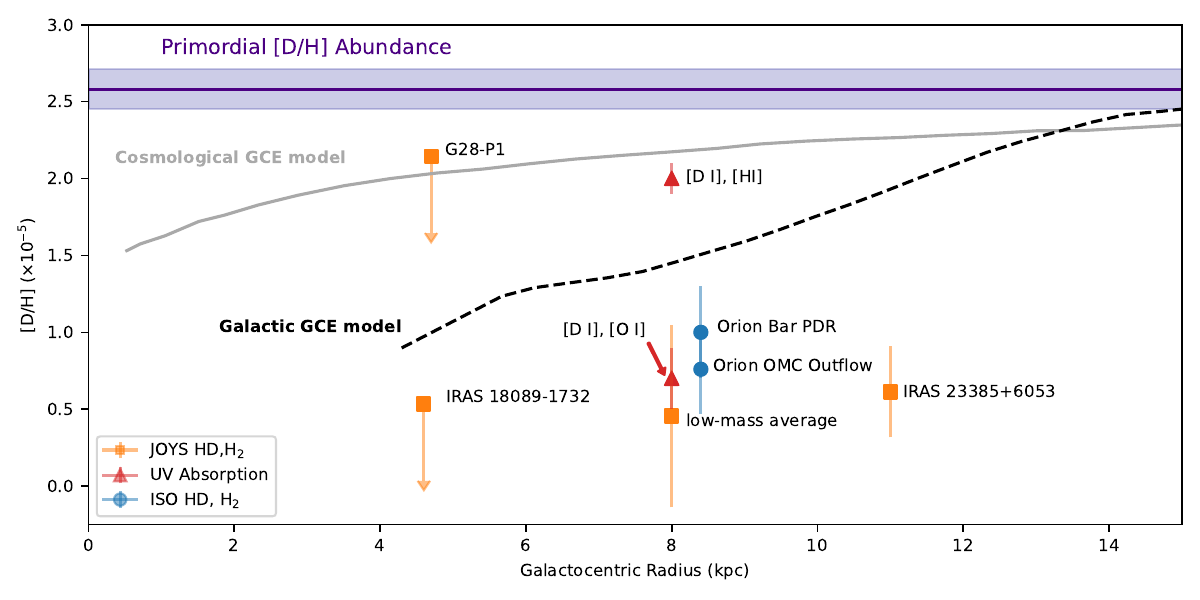}
    \caption{[D/H] as a function of Galactocentric radius based on our HD and H$_2$ rotational line measurements within JOYS (squares), other rotational line values  from ISO (circles) \citep{Wright1999,Bertoldi1999}, and local Galactic disk values from UV absorption lines (triangles) \citep{Prodanovic2010,Hebrard2010}. Our measurements and those for the Orion OMC outflow include an upwards correction factor of 2.54 to [D/H] for chemical conversion of HD to D. We also show Galactic chemical evolution models of our Galaxy from  (black dashed line, reproduced from solid line in Fig. 5 of \cite{Romano2006}) and from a Milky Way type Galaxy in a cosmological zoom-in simulation from (grey line, reproduced from  model 'm12i' in Fig. 5 of \cite{vandeVoort2018}).}
    \label{fig:DtoH_rgc}
\end{figure*}

Galactic chemical evolution models make predictions for the total [D/H] abundance throughout the Galactic disk that can in principle be compared with observations. This is useful for both placing constraints on the GCE models themselves and assessing the potential effects of deuterium depletion on the observed [D/H]. We thus plot in Fig. \ref{fig:DtoH_rgc} as a function of \rgc~the average [D/H] of our low-mass sources, the [D/H] measurement or upper limits for our high-mass sources, [D/H] from previous ISO measurements,  and [D/H] measurements based on UV absorption lines. We overlay the primordial deuterium abundance of \cite{Cyburt2016}, as well as total [D/H] gradients from GCE models selected from \cite{Romano2006} and \cite{vandeVoort2018}. 

If we interpret our [D/H] measurements derived from HD and H$_2$ as representative of \emph{total} [D/H], their values are generally much more similar to the low estimates of the local disk [D/H] of \cite{Hebrard2003} and \cite{Hebrard2010} than the high estimates of \cite{Linsky2006} and \cite{Prodanovic2010}. As noted in Sect. \ref{sec:intro}, the debate on which value represents the present Galactic disk [D/H] abundance within a few kpc of the Sun remains unsettled, and the significance of the inferred depletion onto dust grains and the uncertainties in abundances and GCE effects introduced by using [D/O] as a proxy are both debated \citep{Tosi2010}. However, the molecular line measurements and the low local disk [D/H] are typically a factor of $\sim2-4$ below the predictions of the GCE models. The model of \cite{Romano2006} is based on the GCE code of \cite{Chiappini2001}, and simultaneously satisfies existing chemical abundance constraints while being marginally able to reproduce the depletion-corrected [D/H] of \cite{Prodanovic2010}. \cite{Romano2006} have explored whether GCE models can reproduce the low values of [D/H] in the local Galactic disk of $<1.0 \times 10^{-5}$ from \cite{Hebrard2010} by increasing the efficiency of star formation in the solar neighbourhood, and thus the amount of astration. However, the increased star formation also produces more enrichment of the local Galactic disk with metals, and a resulting metallicity distribution for G and K type stars in the solar neighbourhood with much larger [Fe/H] than is observed. Similar problems in fitting this metallicity distribution occur if the amount of infall onto the Galactic disk of deuterium-rich and metal-poor gas is instead decreased.

It is also interesting to compare the [D/H] predictions from GCE models which are not aimed at specifically reproducing the abundance constraints from our Galaxy. In the context of the evolution of the [D/H] abundance in galaxies with cosmic time, \cite{vandeVoort2018} have performed GCE simulations based on cosmological zoom-in simulations for galaxies forming over a wide range of redshifts and metallicity. The selected model shown in Fig. \ref{fig:DtoH_rgc} shows the total [D/H] gradient for a typical milky way-like galaxy. The high [D/H] across the disk is much more consistent with the local disk [D/H] of \cite{Prodanovic2010} that includes a correction for dust depletion.

GCE models generally predict an increasing total [D/H] with galactocentric radius owing to the increased star formation, and thus astration towards the Galactic center. Our high-mass protostar targets with detections or constraining upper limits on [D/H] are located in the inner and outer Galaxy, and may therefore be useful for measuring this predicted gradient, with the caveat of possible dust depletion.  We detect HD lines only in the the local low-mass protostars and in IRAS 23385+6053 at \rgc=11 kpc, but not in any of our inner Galaxy targets. Taken at face value, the upper limit for IRAS 18089-1732 in particular would be consistent with a [D/H] gradient, however, we can not rule out the possibility that deuterium is more depleted onto dust in the inner Galaxy than in the local disk or outer Galaxy. 

\section{Discussion}
\label{sec:disc}

In summary, our results show large variations in [D/H] between our local low-mass sources, and the derived values of [D/H] are generally a factor of $\sim 2-4$ lower with respect to the predictions of Galactic chemical evolution models. Several of our low-mass sources also show a [D/H] at nearly the primordial value (Fig. \ref{fig:DtoH_scatter}). If depletion is the underlying cause for the variations in [D/H], this would suggest that the total [D/H] is underestimated in many of our objects when derived from the molecular lines alone. Variable dust depletion would resolve the tension with the various GCE models shown in Fig. \ref{fig:DtoH_rgc}, which are generally unable to produce a low [D/H] and satisfy other abundance constraints. However, we do not see any clear trends in [D/H] abundance in comparison with the fluxes of the [S I] and [Fe I] lines, which could trace refractory material in dust grains released in shocks. One possible reason is that deuterium is thought to be depleted by replacing H-C bonds in carbonaceous dust grains or deuterated PAHs, with the latter experiencing different shock conditions, and thus the atomic [S I] and [Fe I] lines are not tracing the solid reservoir into which deuterium is depleted. 

An additional source of uncertainty in our [D/H] measurements is the value of the chemical conversion and non-LTE correction factors derived by \cite{Bertoldi1999}. While they show that the combination of these two correction factors is approximately constant across a range of shock conditions, more detailed modelling of individual spectra and shocks is warranted to see what fraction of our variations in [D/H] can be attributed to differences in the excitation conditions and chemical conversion.

To make progress on measuring total [D/H] with JWST, further work is needed on both the observations and modelling. On the observational front, deeper integrations would provide much better upper limits on the HD column density in quiescent regions of outflows where dust destruction is unlikely to occur. Observations of additional distant targets in the inner and outer Galaxy are needed, especially close to the Galactic center, if GCE models are to be constrained. In addition to protostellar outflows, rotational HD lines have also been detected in the IC 443C supernova remnant \citep{Yuan2012}. Observations of HD emission in supernova remnants may provide a useful probe of [D/H] in higher velocity shock environments where grain destruction may be more complete. Observations with MIRI/MRS cover many rotational lines of H$_2$ and HD, but rovibrational transitions of both are found in the near-infrared accessible with NIRSpec. The additional rovibrational transitions would be useful for determining the shock conditions. Rovibrational transitions of HD have already been detected in PDRs \citep{Peeters2024} with NIRspec. Hints of Infrared emission bands from deuterated PAHs have been detected with JWST \citep{Peeters2004,Peeters2024}, and their interpretation will be important for understanding their role as a reservoir for deuterium depletion.

In terms of modelling, it is particularly important to understand the shock conditions and degree of dust destruction. Shock models for H$_2$ \citep[e.g.][]{Kristensen2023} and forbidden atomic line emission \citep[e.g][]{Hollenbach1989} can be used to constrain the shock density, and velocity, and excitation conditions. \citep{CarattioGaratti2015,CarattioGaratti2024}. These are important inputs for determining the degree of non-LTE excitation and chemical conversion of HD that takes place in the shock positions. 
    
\section{Summary and Conclusions}
\label{sec:conc}

Using JWST/MIRI observations from the JOYS program (PIDs: 1290, 1257) we have presented an analysis of rotational H$_2$ and HD lines in outflows from 5 high-mass and 5 low-mass protostars with the goal of constraining the [D/H] abundance. Our results and conclusions can be summarized as follows:

\begin{itemize}
    \item H$_2$ lines are detected towards all selected sources, but HD lines are only detected towards the brightest positions tracing the outflow shock knots in the low-mass sources, and in the one outer Galaxy (\rgc = 11 kpc) high-mass protostar IRAS 23385+6053. Notably, no HD emission is detected in the inner Galaxy high-mass protostars (\rgc~ $\sim$ 4 kpc) where a higher degree of deuterium astration is expected. 
    \item In resolved low-mass outflows, the morphology and flux of the HD emission is strongly correlated with high-excitation H$_2$ transitions, [S I], and [Fe I] line emission, showing a clear association with bright shock knots within the jet and bow-shocks.
    \item The column density or upper limits of H$_2$ and HD towards these sources has been obtained through an LTE rotation diagram analysis. Additional corrections to the HD column density accounting for non-LTE behaviour and chemical conversion have been applied following \cite{Bertoldi1999}. JWST/MIRI can now provide much more precise constraints on the H$_2$ and HD column density than previous facilities, thanks to the high sensitivity and large number of H$_2$ and HD lines in the mid-IR, which allow the excitation conditions to be more robustly constrained than previously possible. 
    \item The H$_2$ rotation diagrams typically shows two components: a warm ($\sim$  400-900 K) and high column density ($\sim 10^{20}-10^{22}$ cm$^{-2}$) component associated with the wide-angle outflows in the integrated line intensity maps, and a hot ($\sim$ 1000-2000 K) and much lower column density component ($\sim 10^{17}-10^{19}$ cm$^{-2}$) associated with bright knots and bow-shocks. In most cases the H$_2$ shows and ortho-to-para ratio of $\sim 3$. For a handful of positions, an ortho-to-para ratio $<2.5$ and evidence of enhanced excitation for the higher energy H$_2$ transitions is seen.
    \item The HD rotation diagrams are well-described by a single temperature component, and typically show column densities $\sim 10^5$ orders magnitude smaller than the warm H$_2$ component, and excitation temperatures intermediate between the warm and hot H$_2$ ($\sim$ 700 - 1000 K.)
    \item Our derived [D/H] abundances show a factor of $\sim4$ variation between low-mass protostar sources and between different positions within the same outflows. However, the gas compared between low-mass protostars and within a given outflow is expected to have experienced the same degree of astration.  Furthermore, our abundances are a factor of $\sim 2-4$ below the predictions of GCE models. This suggests that we are not probing the total [D/H] in the observed protostars. 
    \item The observed variations may be caused by the significant depletion of D into carbonaceous dust grains and varying degrees of grain destruction. The enhanced [D/H] and gas-phase abundance of Fe in the inner jet relative to the bow-shocks in HH 211 is consistent with this scenario.
    \item Most of our measurements of [D/H] or similar literature measurements using rotational lines suggest [D/H] $\lesssim 1.0\times10^{-5}$ over Galactocentric radii from 4.6 to 11 kpc, which is very difficult to reconcile with existing Galactic chemical evolution models while still satisfying other chemical abundance constraints.
    Depletion of D into dust and its release in shocks would be consistent with the variations in [D/H] between low-mass sources and the typically low values of [D/H] with respect to Galactic chemical evolution models. Further modeling of the effects of the grain destruction in the shocks are needed. 
\end{itemize}

If the total [D/H] abundance can be accurately estimated, the detection of HD in IRAS 23385+6053 at a galactocentric distance of 11 kpc, and the upper limits in the inner Galaxy from IRAS 18089-1732 present an exciting opportunity. Additional observations of HD and H$_2$ in protostellar or supernova shocks when combined with models for the effects of depletion and chemical conversion could for the first time sample [D/H] across the Galactic disk, especially closer to the Galactic center and the outer Galaxy where little is known. This would provide powerful new constraints on models for the formation and chemical evolution of our Galaxy. 

\begin{acknowledgements}
We are grateful to John H. Black for useful discussions on modelling of H$_2$ and HD emission, and to the anonymous referee for their comments on the paper.     

This work is based on observations made with the NASA/ESA/CSA James Webb Space Telescope. The data were obtained from the Mikulski Archive for Space Telescopes at the Space Telescope Science Institute, which is operated by the Association of Universities for Research in Astronomy, Inc., under NASA contract NAS 5-03127 for JWST. These observations are associated with program 1290. 

Astrochemistry in Leiden is supported by funding from the European
Research Council (ERC) under the European Union’s Horizon 2020
research and innovation programme (grant agreement No. 101019751
MOLDISK), by the Netherlands Research School for Astronomy (NOVA),
and by grant TOP-1 614.001.751 from the Dutch
Research Council (NWO).

A.C.G. acknowledges support from PRIN-MUR 2022 20228JPA3A “The path to star and planet formation in the JWST era (PATH)” funded by NextGeneration EU and by INAF-GoG 2022 “NIR-dark Accretion Outbursts in Massive Young stellar objects (NAOMY)” and Large Grant INAF 2022 “YSOs Outflows, Disks and Accretion: towards a global framework for the evolution of
planet forming systems (YODA)”.

P.J.K. acknowledge support from the Science Foundation Ireland/Irish Research Council Pathway program under grant No. 21/PATH-S/9360

\end{acknowledgements}

\bibliographystyle{aa} % style aa.bst
\bibliography{references.bib} % your references Yourfile.bib

\begin{appendix}

\section{Analyzed Spectral Lines}
\label{sec:app_transitions}

We report the H$_2$, HD, and forbidden atomic lines analyzed in this work in Table \ref{tab:transitions}.

\begin{table}[htb]
    \centering
    \caption{Analyzed H$_2$, HD, and forbidden atomic lines.}
    \begin{tabular}{ccccc}
    \toprule\toprule
     Species & Transition  & $\lambda$ ($\mu$m) & $E_\mathrm{up}$ (K) & A$_\mathrm{ul}$ (s$^{-1}$) \\ \midrule
     HD      & (0,0) R(9) & 12.4805              & 6656.3              & $4.97 \times 10^{-5}$ \\
     HD      & (0,0) R(8) & 13.5961              & 5503.5              & $3.87 \times 10^{-5}$ \\
     HD      & (0,0) R(7) & 15.2510             & 4445.2              & $2.88 \times 10^{-5}$ \\
     HD      & (0,0) R(6) & 16.8938             & 3487.4              & $2.03 \times 10^{-5}$ \\
     HD      & (0,0) R(5) & 19.4310             & 2635.8              & $1.33 \times 10^{-5}$ \\
     HD      & (0,0) R(4) & 23.0338             & 1895.3              & $7.91 \times 10^{-6}$ \\ \midrule
     H$_2$     & (0,0) S(8) &  5.0531 & 8677.1 & $3.23 \times  10^{-7}$ \\ 
     H$_2$     & (0,0) S(7) &  5.5111 & 7196.7 & $2.00 \times  10^{-7}$ \\
     H$_2$     & (0,0) S(6) &  6.1086 & 5829.8 & $1.14 \times  10^{-7}$ \\
     H$_2$     & (0,0) S(5) &  6.9095 & 4586.1 & $5.88 \times  10^{-8}$ \\
     H$_2$     & (0,0) S(4) &  8.0250 & 3474.5 & $2.64 \times  10^{-8}$ \\
     H$_2$     & (0,0) S(3) &  9.6649 & 2503.7 & $9.83 \times  10^{-9}$ \\       
     H$_2$     & (0,0) S(2) & 12.2786 & 1681.6 & $2.75 \times  10^{-9}$ \\
     H$_2$     & (0,0) S(1) & 17.0348& 1015.1 & $4.76 \times  10^{-10}$ \\
     \midrule
     {[S I]}      & $^3$P$_1$-$^3$P$_2$       & 25.2490 & 569.83 & $8.27 \times 10^{-9}$ \\
     {[Fe I]}     & a$^5$D$_3$-a$^5$D$_4$     & 24.0423 & 598.43 & $2.51 \times 10^{-3}$ \\
     {[Fe II]}    & a$^4$F$_{7/2}$-a$^4$F$_{9/2}$  & 17.9359 & 3496.42 & $5.86 \times 10^{-3}$\\
    \midrule
    \end{tabular}
    \label{tab:transitions}
\end{table}
%\newpage

\section{Aperture Locations}
\label{sec:app_apertures}

We describe the position of the apertures used for spectral extraction towards each source in Table \ref{tab:apertures}.

\begin{table}[htb]
    \centering
    \caption{Position of our extracted aperture centers in each source. Each aperture is 1\arcsec in radius.}
\begin{tabular}{lrll}
\toprule
Source & Aperture & RA (J200) & Dec (J2000) \\
\midrule
G28-IRS2 & 1 & 18:42:51.89 & -3:59:53.0 \\
G28-P1-A & 1 & 18:42:50.72 & -4:03:16.9 \\
G31-A & 1 & 18:47:34.31 & -1:12:45.0 \\
IRAS 18089-1732 & 1 & 18:11:51.47 & -17:31:24.8 \\
IRAS 23385+6053 & 1 & 23:40:54.80 & +61:10:28.5 \\
BHR71-IRS1 & 1 & 12:01:36.40 & -65:08:51.5 \\
BHR71-IRS1 & 2 & 12:01:36.48 & -65:08:53.5 \\
BHR71-IRS1 & 3 & 12:01:36.36 & -65:08:55.5 \\
BHR71-IRS1 & 4 & 12:01:36.43 & -65:08:57.6 \\
HH 211 & 1 & 3:43:57.34 & 32:00:46.9 \\
HH 211 & 2 & 3:43:57.52 & 32:00:45.6 \\
HH 211 & 3 & 3:43:57.76 & 32:00:44.3 \\
HH 211 & 4 & 3:43:58.26 & 32:00:41.6 \\
HH 211 & 5 & 3:43:58.59 & 32:00:41.5 \\
HH 211 & 6 & 3:43:58.90 & 32:00:35.6 \\
HH 211 & 7 & 3:43:59.39 & 32:00:35.1 \\
HH 211 & 8 & 3:43:59.64 & 32:00:34.1 \\
HH 211 & 9 & 3:43:59.88 & 32:00:34.4 \\
IRAS 4B & 1 & 3:29:12.05 & 31:13:02.1 \\
IRAS 4B & 2 & 3:29:12.09 & 31:13:00.7 \\
L1448-mm & 1 & 3:25:38.82 & 30:44:07.5 \\
L1448-mm & 2 & 3:25:38.74 & 30:44:09.3 \\
L1448-mm & 3 & 3:25:38.67 & 30:44:11.2 \\
L1448-mm & 4 & 3:25:38.57 & 30:44:09.3 \\
Ser-emb 8 (N) & 1 & 18:29:49.34 & 1:16:52.0 \\
Ser-emb 8 (N) & 2 & 18:29:49.19 & 1:16:53.1 \\
\bottomrule
\end{tabular}
    \label{tab:apertures}
\end{table}

\FloatBarrier

\newpage
\section{Integrated Line Intensity Maps}
\label{sec:app_line_maps}

We show below the integrated line intensity maps for all sources in Table \ref{tab:jwst_targets} except IRAS 23385+6053 and HH211, which are shown in Sect. \ref{sec:line_maps}.

\begin{figure}[htb]
    \centering
    \includegraphics[width=\textwidth]{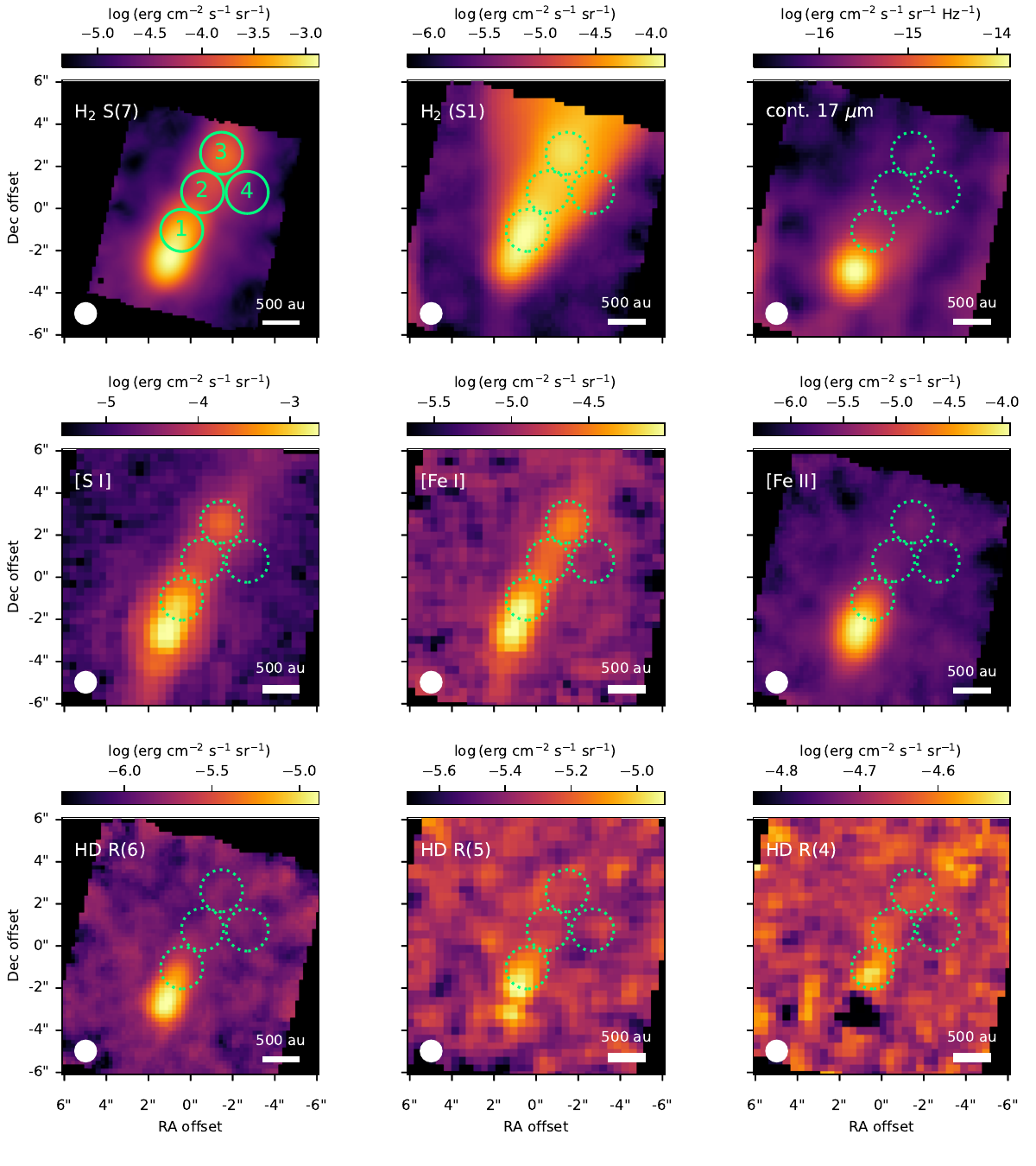}
    \caption{Same as Fig. \ref{fig:moment0_1}, but for L1448-mm.}
    \label{fig:moment0_3}
\end{figure}

\begin{figure*}[htb]
    \centering
    \includegraphics[width=\textwidth]{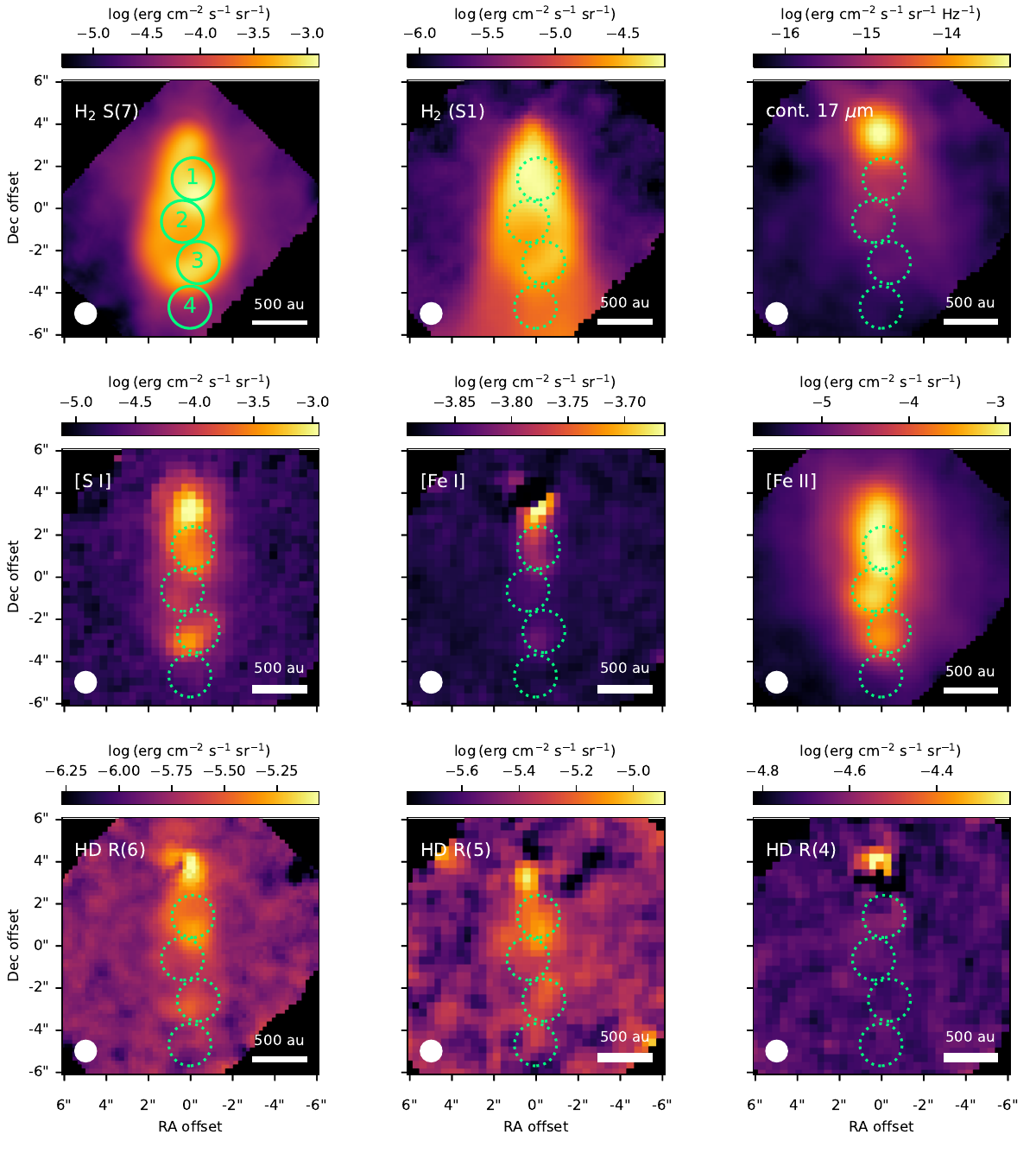}
    \caption{Same as Fig. \ref{fig:moment0_1}, but for BHR71-IRS1.}
    \label{fig:moment0_4}
\end{figure*}

\begin{figure*}[htb]
    \centering
    \includegraphics[width=\textwidth]{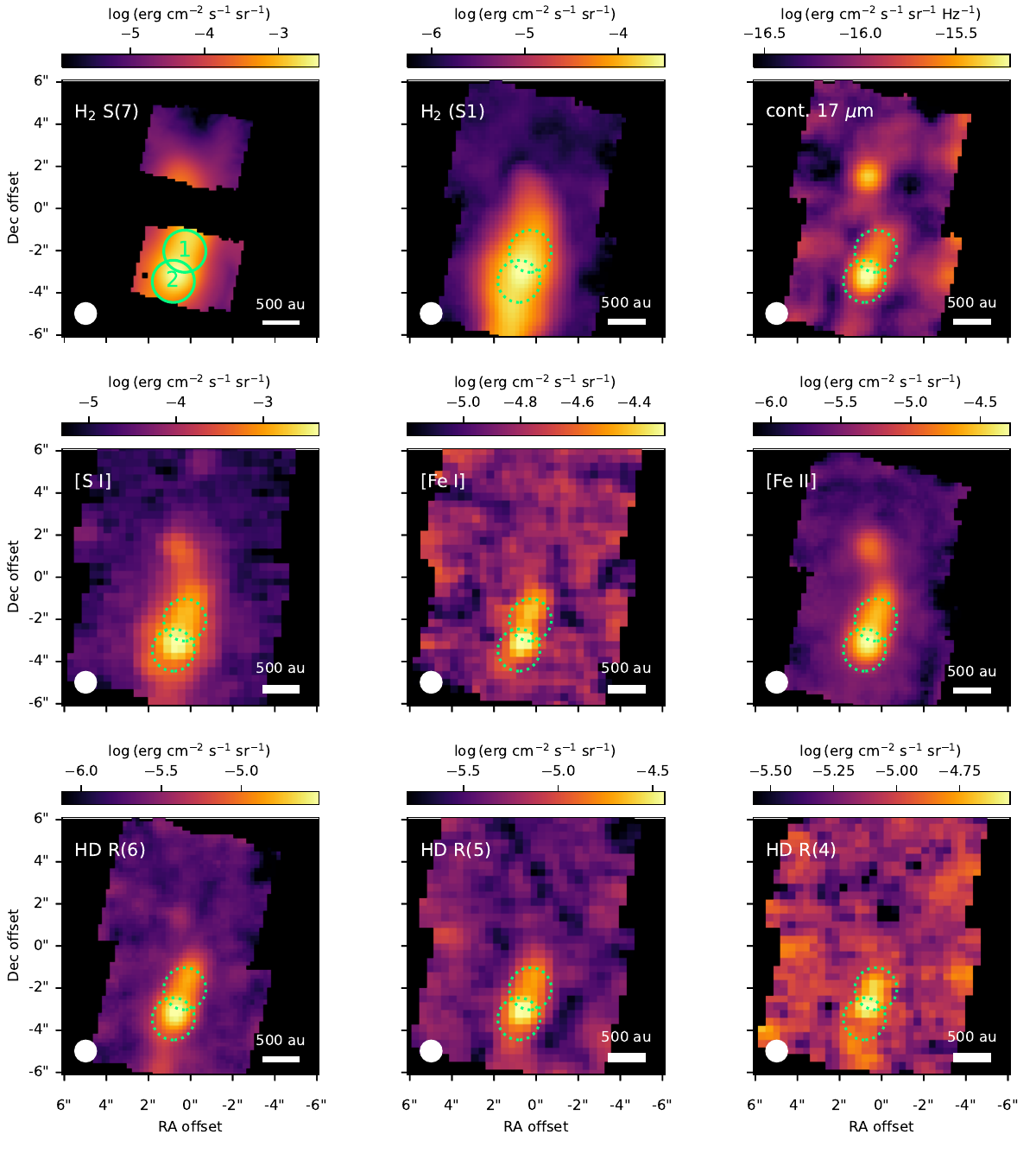}
    \caption{Same as Fig. \ref{fig:moment0_1}, but for IRAS4B.}
    \label{fig:moment0_5}
\end{figure*}

\begin{figure*}[htb]
    \centering
    \includegraphics[width=\textwidth]{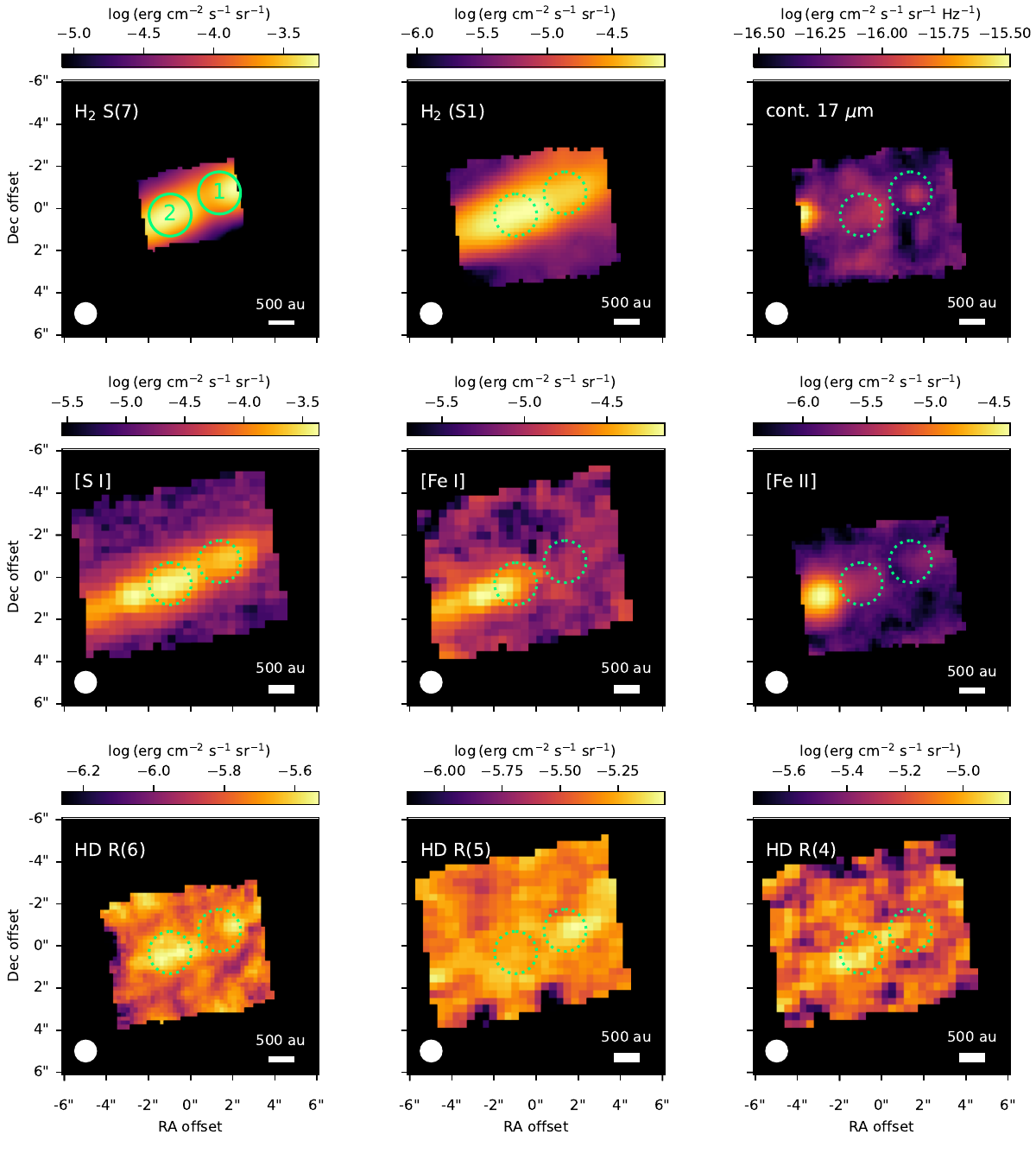}
    \caption{Same as Fig. \ref{fig:moment0_1}, but for Ser-emb8N.}
    \label{fig:moment0_6}
\end{figure*}

\begin{figure*}[htb]
    \centering
    \includegraphics[width=\textwidth]{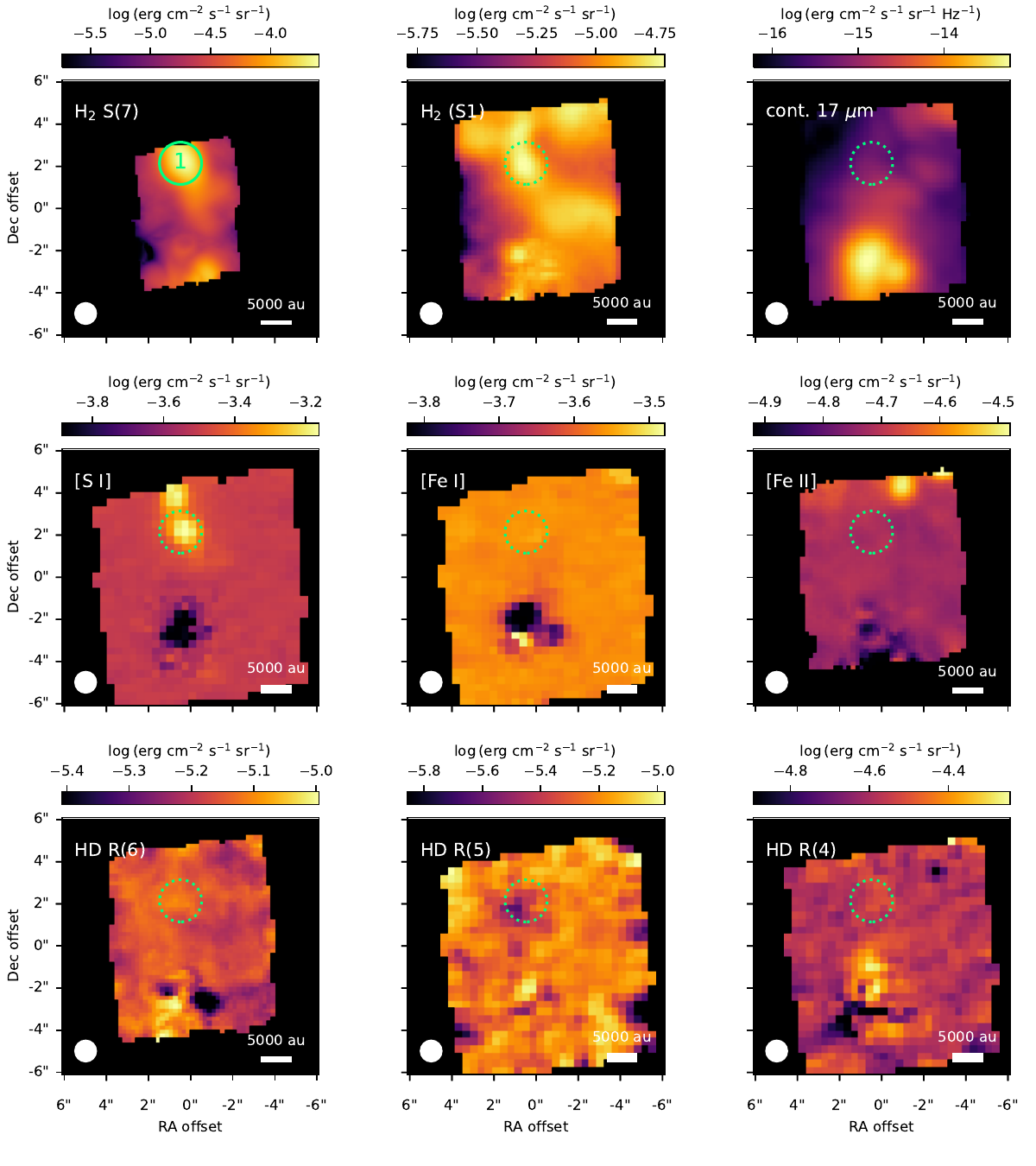}
    \caption{Same as Fig. \ref{fig:moment0_1}, but for IRAS 18089-1732.}
    \label{fig:moment0_7}
\end{figure*}

\begin{figure*}[htb]
    \centering
    \includegraphics[width=\textwidth]{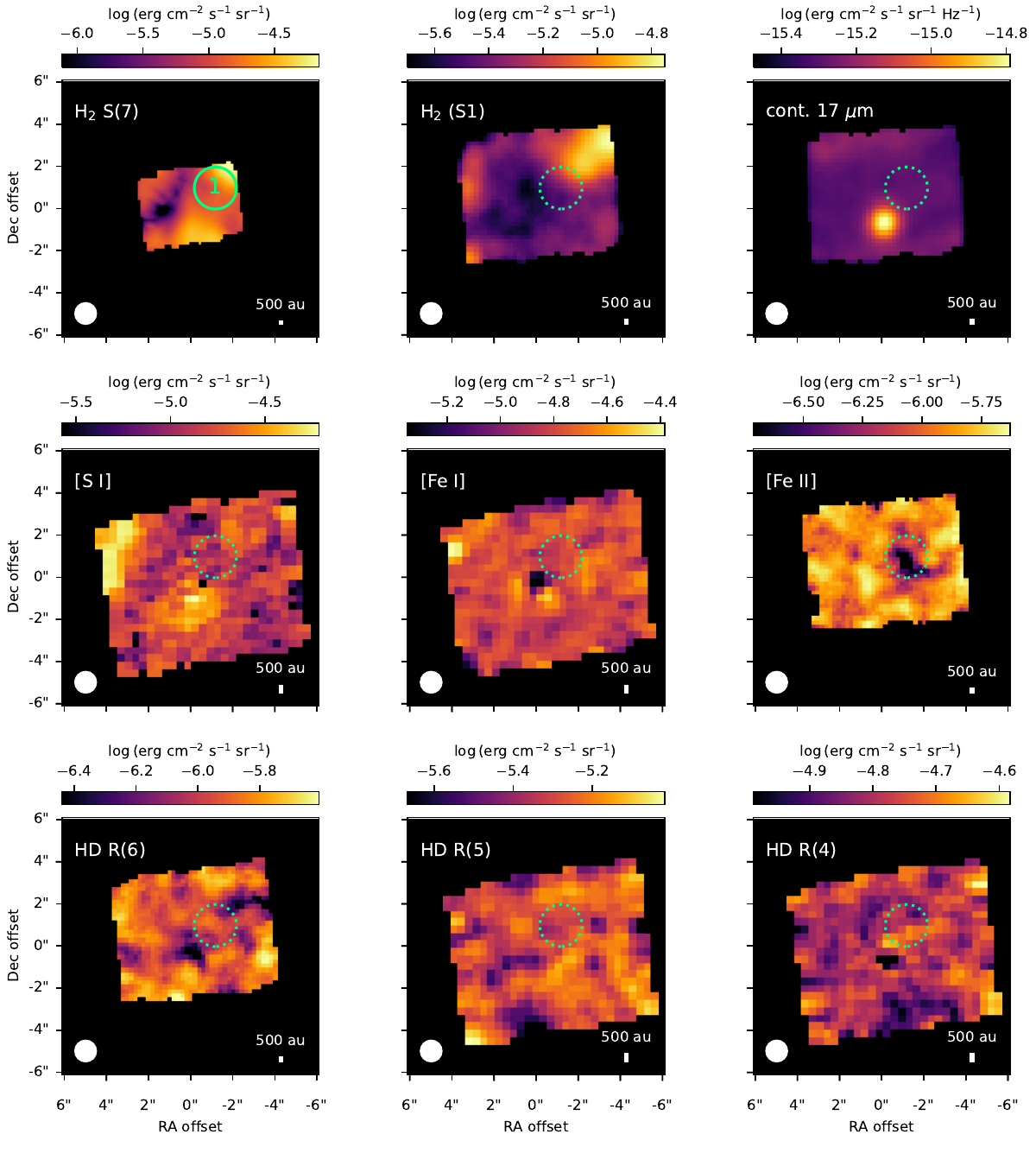}
    \caption{Same as Fig. \ref{fig:moment0_1}, but for G28-IRS2.}
    \label{fig:moment0_8}
\end{figure*}

\begin{figure*}[htb]
    \centering
    \includegraphics[width=\textwidth]{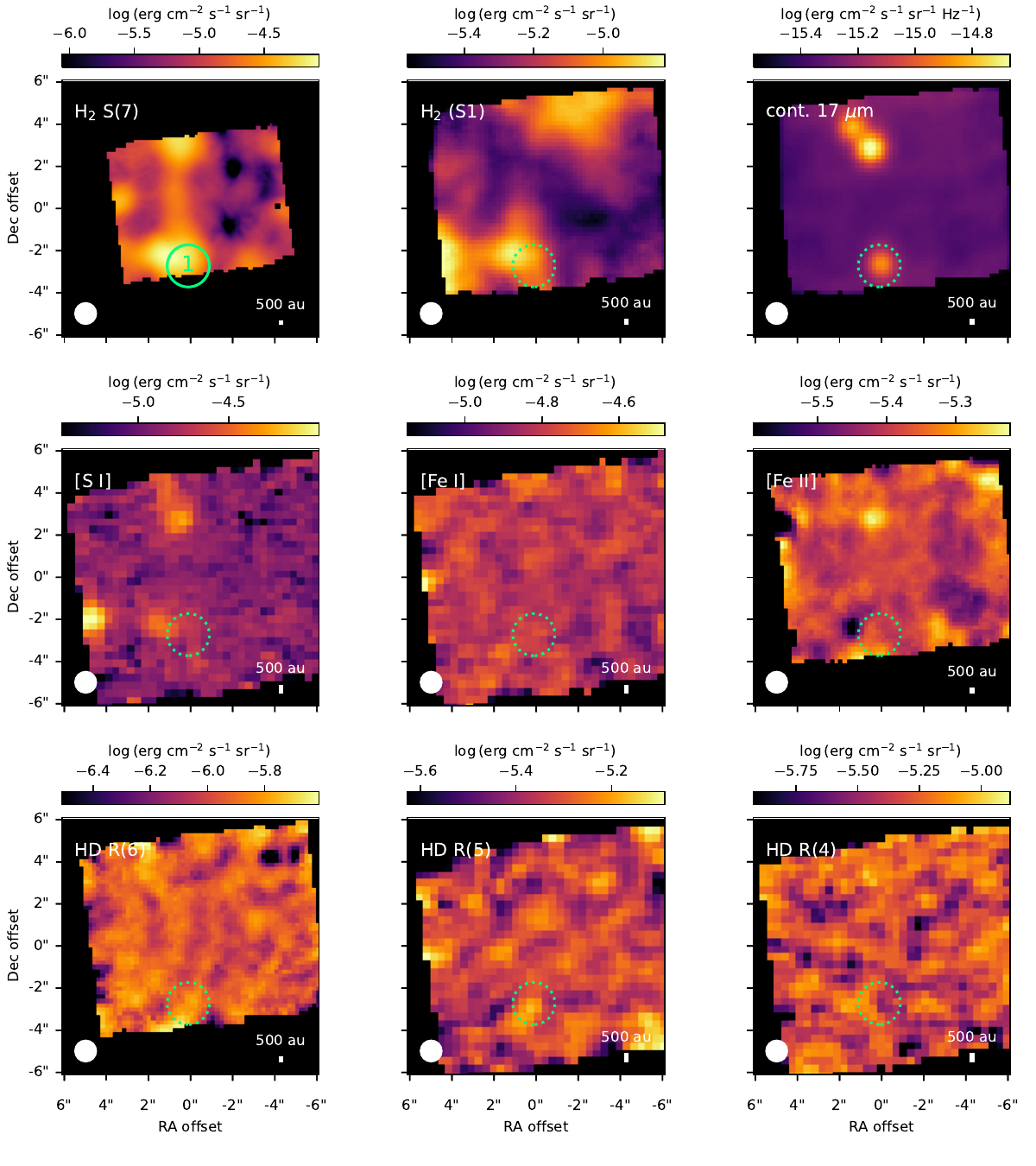}
    \caption{Same as Fig. \ref{fig:moment0_1}, but for G28-P1.}
    \label{fig:moment0_9}
\end{figure*}

\begin{figure*}[htb]
    \centering
    \includegraphics[width=\textwidth]{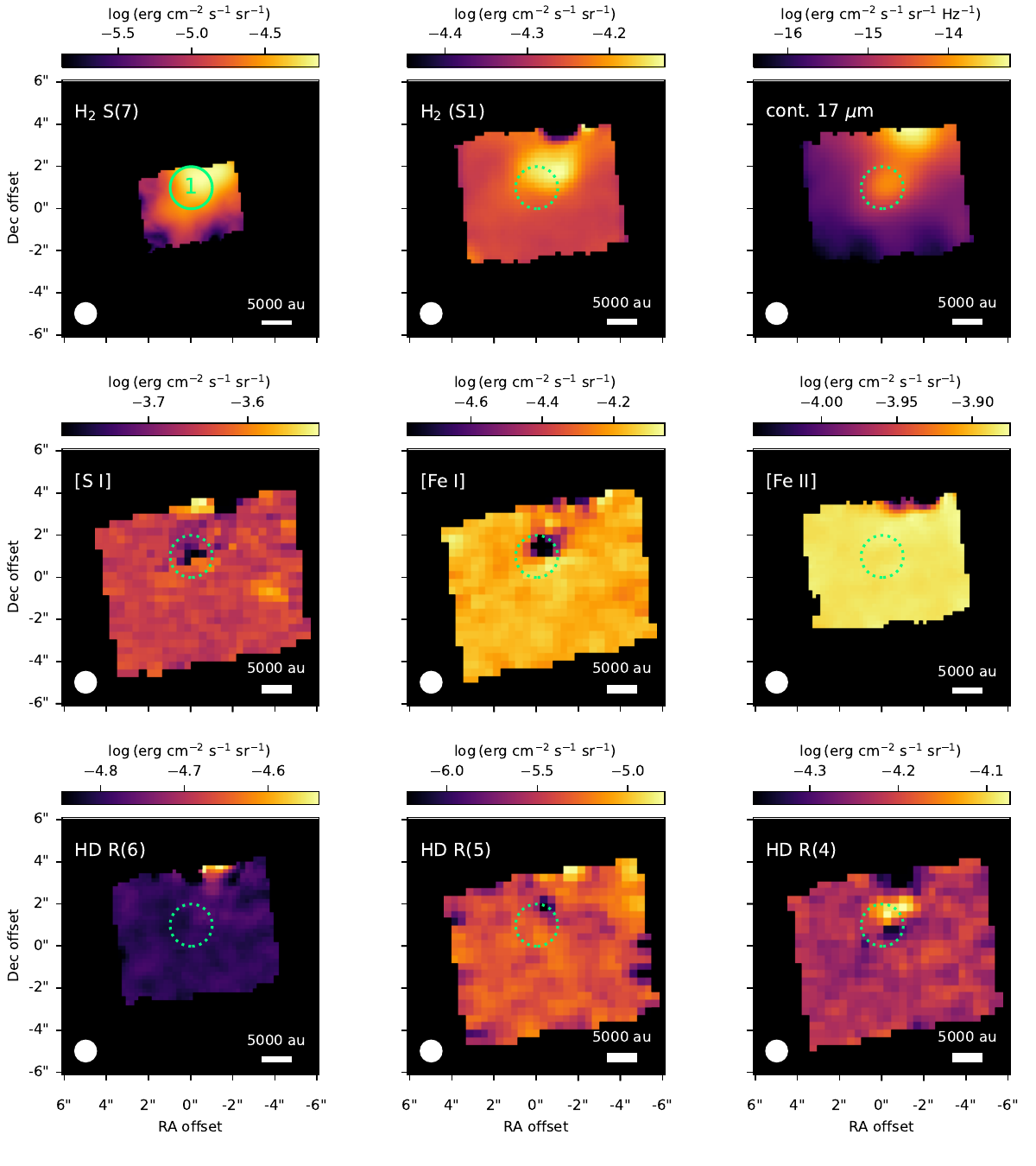}
    \caption{Same as Fig. \ref{fig:moment0_1}, but for G31.}
    \label{fig:moment0_10}
\end{figure*}

\FloatBarrier

\clearpage
\section{Example H$_2$ and HD line detections}
\label{sec:app_iras23385_line_det}

\begin{figure}[h]
    \centering
    \includegraphics[width=\textwidth]{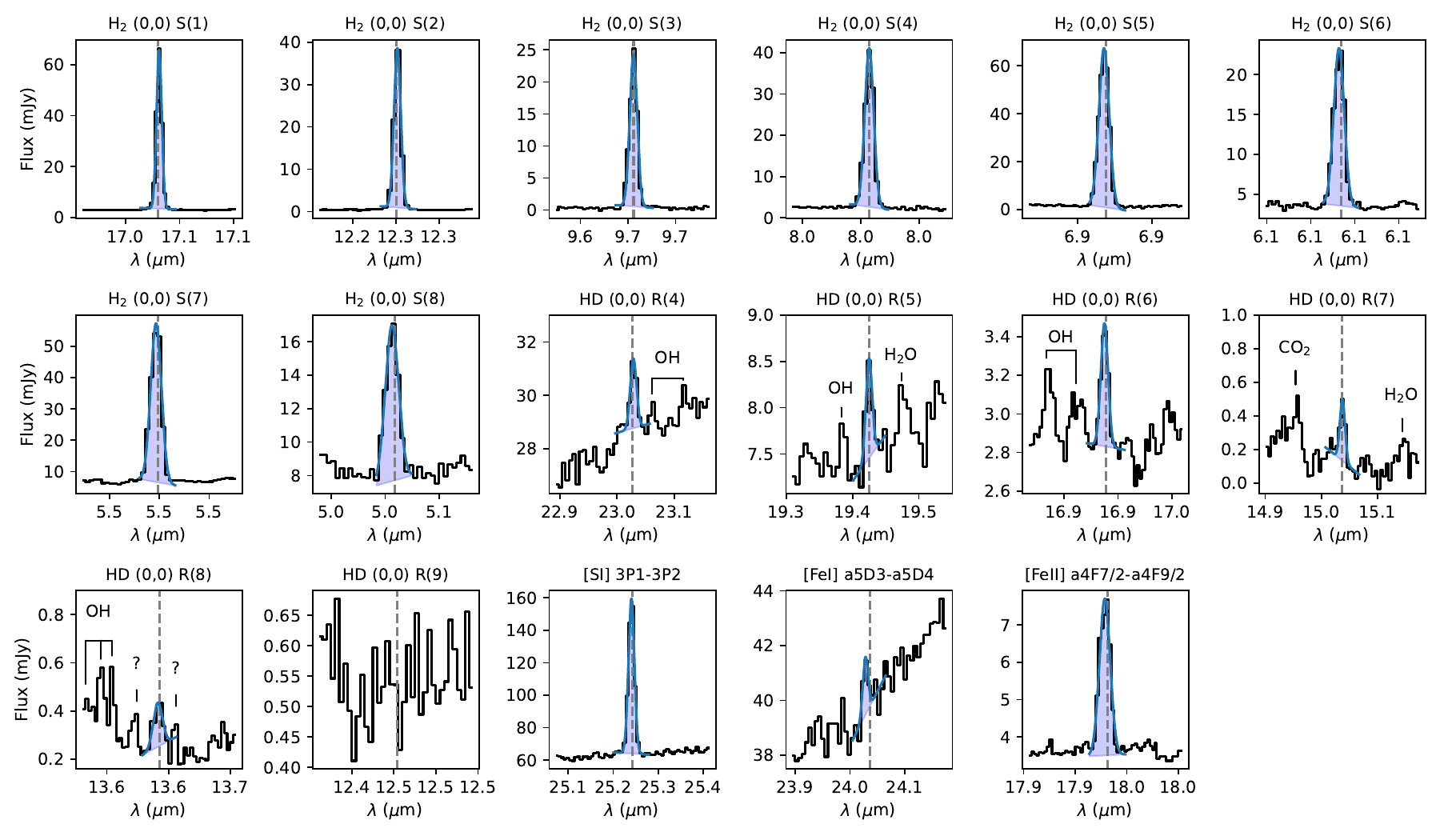}
    \caption{Same as Fig. \ref{fig:HH211_line_det}, but for IRAS 23385+6053 aperture 1.}
    \label{fig:IRAS23385_line_det}
\end{figure}

In Figure \ref{fig:IRAS23385_line_det}, we show additional examples of detected H$_2$, HD, and selected atomic lines.

\clearpage

\section{Line Fluxes}
\label{sec:app_line_fluxes}
In Tables \ref{tab:h2_line_fluxes} to \ref{tab:atomic_line_fluxes} we provide the H$_2$, HD, and atomic line fluxes or upper limits for every aperture in Table \ref{tab:apertures}.

\begin{table}[htb]
    \tiny
    \caption{H$_2$ Line fluxes and upper limits, units of $10^{-15}$ erg cm$^{-2}$ s$^{-1}$. No extinction correction is applied.}
    \label{tab:h2_line_fluxes}
\begin{tabular}{lrllllllll}
\toprule\toprule
         Source &  Ap. &       H$_2$ S(1) &       H$_2$ S(2) &       H$_2$ S(3) &       H$_2$ S(4) &        H$_2$ S(5) &       H$_2$ S(6) &        H$_2$ S(7) &       H$_2$ S(8) \\
\midrule
G28-IRS2 & 1 & $0.96 \pm 0.02$ & $0.36 \pm 0.01$ & $0.19 \pm 0.03$ & $0.57 \pm 0.01$ & $0.53 \pm 0.04$ & $< 0.09$ & $1.70 \pm 0.10$ & $0.28 \pm 0.03$ \\
G28-P1-A & 1 & $1.04 \pm 0.03$ & $0.65 \pm 0.02$ & $0.28 \pm 0.02$ & $1.38 \pm 0.04$ & $1.93 \pm 0.04$ & $0.48 \pm 0.03$ & $2.91 \pm 0.06$ & $0.68 \pm 0.03$ \\
G31-A & 1 & $1.66 \pm 0.05$ & $1.64 \pm 0.02$ & $0.09 \pm 0.02$ & $2.47 \pm 0.06$ & $2.38 \pm 0.08$ & $0.52 \pm 0.02$ & $3.40 \pm 0.07$ & $0.63 \pm 0.04$ \\
IRAS 18089-1732 & 1 & $1.52 \pm 0.05$ & $1.58 \pm 0.02$ & $0.18 \pm 0.01$ & $5.38 \pm 0.07$ & $7.93 \pm 0.04$ & $1.96 \pm 0.02$ & $10.72 \pm 0.05$ & $2.11 \pm 0.03$ \\
IRAS 23385+6053 & 1 & $3.92 \pm 0.07$ & $3.68 \pm 0.12$ & $2.99 \pm 0.21$ & $6.79 \pm 0.39$ & $14.45 \pm 1.02$ & $4.71 \pm 0.42$ & $13.45 \pm 1.31$ & $3.40 \pm 0.51$ \\
BHR71-IRS1 & 1 & $3.95 \pm 0.02$ & $5.51 \pm 0.12$ & $10.50 \pm 0.06$ & $23.49 \pm 0.27$ & $55.30 \pm 0.57$ & $17.15 \pm 0.16$ & $57.26 \pm 0.93$ & $13.94 \pm 0.27$ \\
BHR71-IRS1 & 2 & $2.47 \pm 0.03$ & $3.78 \pm 0.11$ & $12.97 \pm 0.10$ & $14.02 \pm 0.17$ & $40.16 \pm 0.53$ & $12.96 \pm 0.13$ & $42.25 \pm 0.33$ & $11.77 \pm 0.15$ \\
BHR71-IRS1 & 3 & $2.12 \pm 0.02$ & $3.13 \pm 0.06$ & $15.86 \pm 0.25$ & $12.69 \pm 0.34$ & $39.33 \pm 1.02$ & $13.41 \pm 0.32$ & $42.37 \pm 1.32$ & $14.53 \pm 0.32$ \\
BHR71-IRS1 & 4 & $1.08 \pm 0.01$ & $1.11 \pm 0.02$ & $4.02 \pm 0.05$ & $2.22 \pm 0.04$ & $5.64 \pm 0.11$ & $1.61 \pm 0.08$ & $5.31 \pm 0.12$ & $1.47 \pm 0.06$ \\
HH 211 & 1 & $6.62 \pm 0.18$ & $7.57 \pm 0.29$ & $26.67 \pm 0.33$ & $17.26 \pm 0.16$ & $41.01 \pm 0.89$ & $9.85 \pm 0.14$ & $21.27 \pm 0.19$ & $4.47 \pm 0.08$ \\
HH 211 & 2 & $6.89 \pm 0.15$ & $6.44 \pm 0.04$ & $21.32 \pm 0.38$ & $10.29 \pm 0.10$ & $21.97 \pm 0.69$ & $4.98 \pm 0.07$ & $9.24 \pm 0.08$ & $1.98 \pm 0.11$ \\
HH 211 & 3 & $7.54 \pm 0.16$ & $6.98 \pm 0.02$ & $25.00 \pm 0.36$ & $12.59 \pm 0.13$ & $27.77 \pm 0.58$ & $6.25 \pm 0.14$ & $11.83 \pm 0.12$ & $2.27 \pm 0.08$ \\
HH 211 & 4 & $2.71 \pm 0.08$ & $2.95 \pm 0.04$ & $8.69 \pm 0.13$ & $3.96 \pm 0.07$ & $8.29 \pm 0.11$ & $1.84 \pm 0.05$ & $4.00 \pm 0.07$ & $1.01 \pm 0.06$ \\
HH 211 & 5 & $1.83 \pm 0.04$ & $3.46 \pm 0.06$ & $7.27 \pm 0.07$ & $5.22 \pm 0.05$ & $8.76 \pm 0.13$ & $2.35 \pm 0.05$ & $4.74 \pm 0.10$ & $1.22 \pm 0.07$ \\
HH 211 & 6 & $1.72 \pm 0.08$ & $2.15 \pm 0.03$ & $5.88 \pm 0.06$ & $2.61 \pm 0.06$ & $5.08 \pm 0.17$ & $1.29 \pm 0.04$ & $2.75 \pm 0.12$ & $0.50 \pm 0.09$ \\
HH 211 & 7 & $13.69 \pm 0.46$ & $45.55 \pm 0.32$ & $63.86 \pm 1.79$ & $67.14 \pm 0.70$ & $120.38 \pm 1.20$ & $46.32 \pm 1.06$ & $98.16 \pm 0.92$ & $30.25 \pm 0.72$ \\
HH 211 & 8 & $6.44 \pm 0.19$ & $12.25 \pm 0.13$ & $38.03 \pm 0.87$ & $28.68 \pm 0.40$ & $77.85 \pm 1.32$ & $24.82 \pm 0.57$ & $65.23 \pm 0.59$ & $20.78 \pm 0.67$ \\
HH 211 & 9 & $4.72 \pm 0.14$ & $12.19 \pm 0.22$ & $27.32 \pm 0.52$ & $26.92 \pm 0.27$ & $66.47 \pm 1.59$ & $28.50 \pm 0.37$ & $66.35 \pm 0.50$ & $26.72 \pm 1.22$ \\
IRAS 4B & 1 & $11.36 \pm 0.42$ & $18.70 \pm 0.22$ & $16.27 \pm 0.09$ & $52.84 \pm 0.75$ & $91.97 \pm 1.56$ & $36.45 \pm 0.69$ & $103.99 \pm 0.63$ & $33.24 \pm 0.68$ \\
IRAS 4B & 2 & $14.88 \pm 0.49$ & $24.38 \pm 0.23$ & $22.95 \pm 0.13$ & $62.85 \pm 0.81$ & $99.36 \pm 1.01$ & $45.21 \pm 0.67$ & $113.75 \pm 1.15$ & $34.98 \pm 0.64$ \\
L1448-mm & 1 & $7.26 \pm 0.09$ & $6.93 \pm 0.04$ & $25.01 \pm 0.17$ & $20.36 \pm 0.13$ & $47.42 \pm 0.57$ & $11.82 \pm 0.07$ & $35.52 \pm 0.39$ & $7.42 \pm 0.12$ \\
L1448-mm & 2 & $4.78 \pm 0.03$ & $4.43 \pm 0.05$ & $15.53 \pm 0.11$ & $8.45 \pm 0.12$ & $17.10 \pm 0.31$ & $3.54 \pm 0.05$ & $7.55 \pm 0.11$ & $1.39 \pm 0.07$ \\
L1448-mm & 3 & $5.35 \pm 0.04$ & $5.63 \pm 0.06$ & $19.72 \pm 0.16$ & $10.15 \pm 0.12$ & $21.47 \pm 0.28$ & $4.54 \pm 0.05$ & $9.98 \pm 0.13$ & $2.04 \pm 0.12$ \\
L1448-mm & 4 & $1.77 \pm 0.03$ & $1.70 \pm 0.01$ & $5.11 \pm 0.05$ & $2.68 \pm 0.06$ & $4.53 \pm 0.07$ & $0.84 \pm 0.04$ & $1.93 \pm 0.10$ & $0.40 \pm 0.03$ \\
Ser-emb 8 (N) & 1 & $2.96 \pm 0.02$ & $6.71 \pm 0.05$ & $6.29 \pm 0.06$ & $17.19 \pm 0.20$ & $28.85 \pm 0.54$ & $10.27 \pm 0.23$ & $22.88 \pm 0.19$ & $6.43 \pm 0.10$ \\
Ser-emb 8 (N) & 2 & $4.45 \pm 0.02$ & $9.13 \pm 0.05$ & $10.07 \pm 0.11$ & $21.46 \pm 0.20$ & $42.65 \pm 0.36$ & $11.80 \pm 0.09$ & $28.69 \pm 0.30$ & $7.16 \pm 0.23$ \\
\midrule
\end{tabular}
\end{table}

\begin{table*}[htb]
    \centering
    \small
    \caption{HD Line fluxes and upper limits, units of $10^{-15}$ erg cm$^{-2}$ s$^{-1}$. No extinction correction is applied.}
    \label{tab:hd_line_fluxes}
\begin{tabular}{lrllllll}
\toprule\toprule
         Source &  Aperture &         HD R(4) &         HD R(5) &         HD R(6) &         HD R(7) &         HD R(8) &         HD R(9) \\
\midrule
G28-IRS2 & 1 & $< 0.12$ & $< 0.04$ & $< 0.02$ & $< 0.02$ & $< 0.02$ & $< 0.02$ \\
G28-P1-A & 1 & $< 0.09$ & $< 0.03$ & $< 0.02$ & $< 0.03$ & $< 0.03$ & $< 0.02$ \\
G31-A & 1 & $< 0.92$ & $< 0.15$ & $< 0.03$ & $< 0.25$ & $< 0.04$ & $< 0.03$ \\
IRAS 18089-1732 & 1 & $< 0.16$ & $< 0.07$ & $< 0.02$ & $< 0.10$ & $< 0.05$ & $< 0.03$ \\
IRAS 23385+6053 & 1 & $0.20 \pm 0.03$ & $0.10 \pm 0.02$ & $0.06 \pm 0.00$ & $0.03 \pm 0.00$ & $0.03 \pm 0.01$ & $< 0.01$ \\
BHR71-IRS1 & 1 & $0.29 \pm 0.09$ & $0.21 \pm 0.05$ & $0.18 \pm 0.01$ & $0.07 \pm 0.00$ & $0.05 \pm 0.01$ & $0.05 \pm 0.01$ \\
BHR71-IRS1 & 2 & $0.16 \pm 0.03$ & $0.10 \pm 0.02$ & $0.09 \pm 0.01$ & $0.06 \pm 0.01$ & $0.03 \pm 0.00$ & $0.03 \pm 0.01$ \\
BHR71-IRS1 & 3 & $< 0.10$ & $0.08 \pm 0.02$ & $0.08 \pm 0.01$ & $0.05 \pm 0.00$ & $0.03 \pm 0.00$ & $0.07 \pm 0.02$ \\
BHR71-IRS1 & 4 & $< 0.07$ & $< 0.05$ & $< 0.02$ & $< 0.03$ & $< 0.02$ & $< 0.02$ \\
HH 211 & 1 & $0.22 \pm 0.03$ & $0.13 \pm 0.03$ & $0.08 \pm 0.01$ & $0.06 \pm 0.01$ & $< 0.03$ & $< 0.02$ \\
HH 211 & 2 & $0.18 \pm 0.05$ & $0.09 \pm 0.02$ & $0.03 \pm 0.01$ & $< 0.03$ & $< 0.03$ & $< 0.02$ \\
HH 211 & 3 & $0.22 \pm 0.04$ & $0.10 \pm 0.01$ & $0.05 \pm 0.01$ & $0.01 \pm 0.01$ & $< 0.03$ & $< 0.03$ \\
HH 211 & 4 & $< 0.11$ & $< 0.05$ & $< 0.02$ & $< 0.03$ & $< 0.03$ & $< 0.02$ \\
HH 211 & 5 & $< 0.13$ & $< 0.08$ & $< 0.02$ & $< 0.02$ & $< 0.02$ & $< 0.02$ \\
HH 211 & 6 & $< 0.12$ & $< 0.08$ & $< 0.03$ & $< 0.05$ & $< 0.04$ & $< 0.03$ \\
HH 211 & 7 & $0.93 \pm 0.10$ & $0.61 \pm 0.08$ & $0.37 \pm 0.04$ & $0.21 \pm 0.02$ & $0.13 \pm 0.01$ & $0.10 \pm 0.02$ \\
HH 211 & 8 & $0.32 \pm 0.04$ & $0.29 \pm 0.07$ & $0.20 \pm 0.02$ & $0.13 \pm 0.01$ & $0.09 \pm 0.01$ & $0.09 \pm 0.01$ \\
HH 211 & 9 & $0.19 \pm 0.11$ & $0.85 \pm 0.69$ & $0.25 \pm 0.05$ & $0.20 \pm 0.02$ & $0.12 \pm 0.01$ & $0.09 \pm 0.02$ \\
IRAS 4B & 1 & $0.83 \pm 0.29$ & $0.60 \pm 0.20$ & $0.74 \pm 0.10$ & $0.26 \pm 0.02$ & $0.29 \pm 0.05$ & $0.19 \pm 0.02$ \\
IRAS 4B & 2 & $0.89 \pm 0.39$ & $0.66 \pm 0.25$ & $0.97 \pm 0.16$ & $0.39 \pm 0.03$ & $0.34 \pm 0.08$ & $0.25 \pm 0.03$ \\
L1448-mm & 1 & $0.39 \pm 0.02$ & $0.22 \pm 0.04$ & $0.15 \pm 0.01$ & $0.10 \pm 0.00$ & $0.03 \pm 0.01$ & $0.03 \pm 0.01$ \\
L1448-mm & 2 & $0.12 \pm 0.02$ & $0.05 \pm 0.02$ & $0.02 \pm 0.00$ & $< 0.02$ & $< 0.01$ & $< 0.01$ \\
L1448-mm & 3 & $0.13 \pm 0.02$ & $0.04 \pm 0.02$ & $0.02 \pm 0.01$ & $< 0.02$ & $< 0.01$ & $< 0.01$ \\
L1448-mm & 4 & $< 0.09$ & $< 0.04$ & $< 0.02$ & $< 0.02$ & $< 0.01$ & $< 0.01$ \\
Ser-emb 8 (N) & 1 & $0.20 \pm 0.07$ & $0.15 \pm 0.04$ & $0.09 \pm 0.01$ & $< 0.03$ & $< 0.02$ & $< 0.02$ \\
Ser-emb 8 (N) & 2 & $0.33 \pm 0.05$ & $0.14 \pm 0.02$ & $0.12 \pm 0.01$ & $0.04 \pm 0.01$ & $0.02 \pm 0.01$ & $0.03 \pm 0.01$ \\
\midrule
\end{tabular}
\end{table*}

\begin{table*}[htb]
    \centering
    \small
    \caption{Atomic Line fluxes and upper limits, units of $10^{-15}$ erg cm$^{-2}$ s$^{-1}$. No extinction correction is applied.}
    \label{tab:atomic_line_fluxes}
\begin{tabular}{lrlll}
\toprule\toprule
         Source &  Aperture &      [SI] 3P1-3P2 & [FeI] a5D3-a5D4 & [FeII] a4F7/2-a4F9/2 \\
\midrule
G28-IRS2 & 1 & $0.35 \pm 0.08$ & $< 0.16$ & $< 0.02$ \\
G28-P1-A & 1 & $0.65 \pm 0.08$ & $< 0.13$ & $< 0.02$ \\
G31-A & 1 & $< 3.16$ & $< 1.32$ & $< 0.04$ \\
IRAS 18089-1732 & 1 & $14.29 \pm 0.52$ & $< 0.29$ & $< 0.03$ \\
IRAS 23385+6053 & 1 & $7.47 \pm 0.13$ & $0.13 \pm 0.05$ & $0.51 \pm 0.06$ \\
BHR71-IRS1 & 1 & $19.43 \pm 0.34$ & $1.69 \pm 0.14$ & $81.71 \pm 3.30$ \\
BHR71-IRS1 & 2 & $6.59 \pm 0.09$ & $0.68 \pm 0.06$ & $55.90 \pm 1.22$ \\
BHR71-IRS1 & 3 & $11.82 \pm 0.28$ & $0.56 \pm 0.04$ & $12.89 \pm 0.13$ \\
BHR71-IRS1 & 4 & $1.34 \pm 0.05$ & $< 0.12$ & $0.91 \pm 0.00$ \\
HH 211 & 1 & $13.38 \pm 0.75$ & $2.31 \pm 0.15$ & $0.12 \pm 0.01$ \\
HH 211 & 2 & $4.47 \pm 0.24$ & $0.64 \pm 0.07$ & $0.03 \pm 0.01$ \\
HH 211 & 3 & $7.78 \pm 0.44$ & $1.55 \pm 0.06$ & $0.08 \pm 0.01$ \\
HH 211 & 4 & $0.66 \pm 0.07$ & $< 0.17$ & $< 0.03$ \\
HH 211 & 5 & $< 0.16$ & $< 0.17$ & $< 0.02$ \\
HH 211 & 6 & $< 0.23$ & $< 0.22$ & $< 0.03$ \\
HH 211 & 7 & $23.76 \pm 1.13$ & $0.28 \pm 0.06$ & $7.29 \pm 0.04$ \\
HH 211 & 8 & $15.72 \pm 0.61$ & $< 0.25$ & $1.47 \pm 0.02$ \\
HH 211 & 9 & $20.61 \pm 0.81$ & $< 0.26$ & $2.66 \pm 0.06$ \\
IRAS 4B & 1 & $100.04 \pm 3.48$ & $1.33 \pm 0.10$ & $1.38 \pm 0.05$ \\
IRAS 4B & 2 & $142.39 \pm 10.32$ & $1.31 \pm 0.13$ & $1.46 \pm 0.06$ \\
L1448-mm & 1 & $41.88 \pm 1.58$ & $3.15 \pm 0.18$ & $1.30 \pm 0.03$ \\
L1448-mm & 2 & $8.16 \pm 0.36$ & $1.24 \pm 0.07$ & $0.07 \pm 0.00$ \\
L1448-mm & 3 & $9.81 \pm 0.59$ & $1.42 \pm 0.07$ & $0.07 \pm 0.01$ \\
L1448-mm & 4 & $1.51 \pm 0.09$ & $0.18 \pm 0.04$ & $< 0.03$ \\
Ser-emb 8 (N) & 1 & $11.21 \pm 0.44$ & $0.23 \pm 0.05$ & $0.05 \pm 0.01$ \\
Ser-emb 8 (N) & 2 & $20.63 \pm 0.89$ & $1.95 \pm 0.14$ & $0.19 \pm 0.01$ \\
\midrule
\end{tabular}
\end{table*}

\clearpage
\section{Rotation diagram extinction and ortho-to-para ratio correction}
\label{sec:ext_corr_app}

\begin{figure*}[!b]
    \centering
    \includegraphics[width=0.8\textwidth]{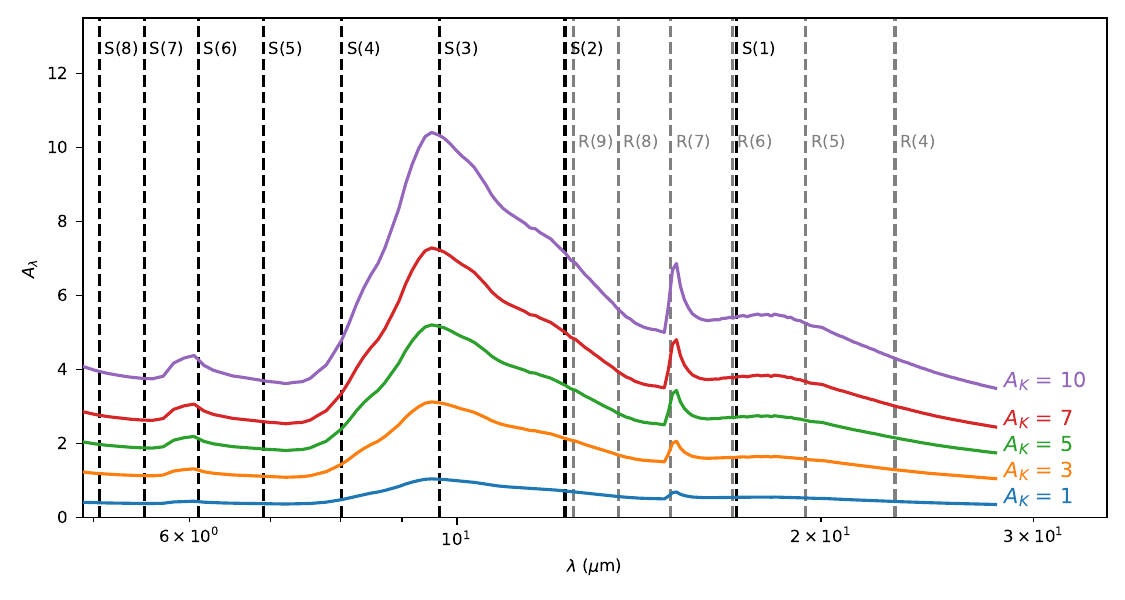}
    \caption{The KP5 \citep{Pontoppidan2024} extinction curve for various values of $A_K$ (solid lines) overlaid on the wavelengths of the H$_2$ and HD rotational transitions (dashed black and grey lines, respectively) analyzed in this work.}
    \label{fig:ext_curve_and_lines}
\end{figure*}

\begin{figure}[htb]
    \centering
    \includegraphics[scale=0.83]{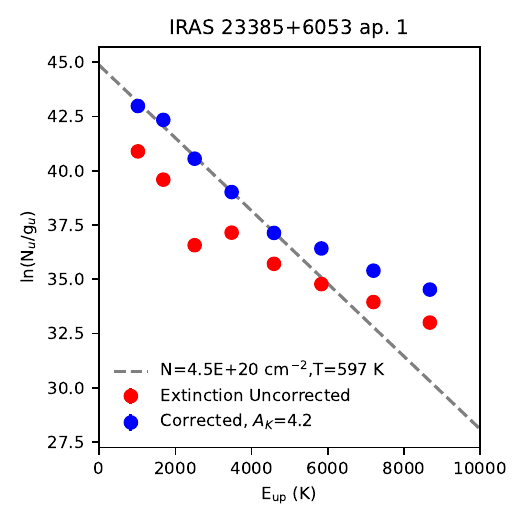}
    \includegraphics[scale=0.83]{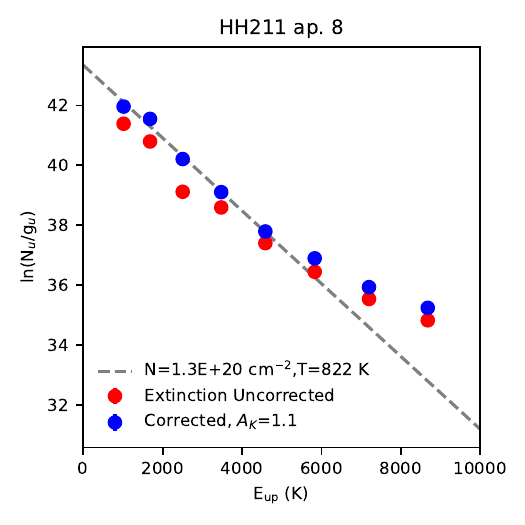}
    \caption{Rotation diagram for IRAS 23385+6053 aperture 1 and HH 211  aperture 8 with a fit to the S(1) to S(4) transitions to determine the extinction $A_K$. The uncorrected intensities are shown as red points, while the extinction corrected values are shown in blue. The dashed grey line shows the best fit of a single temperature component to the S(1) to S(4) transitions.}
    \label{fig:ext_corr_example}
\end{figure}

\begin{figure}[htb]
    \centering
    \includegraphics[scale=0.83]{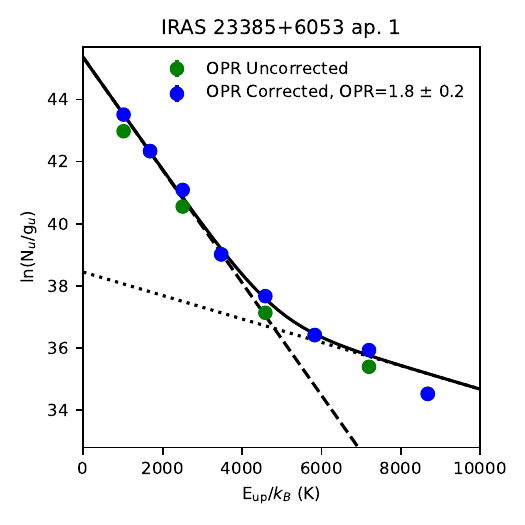}
    \includegraphics[scale=0.83]{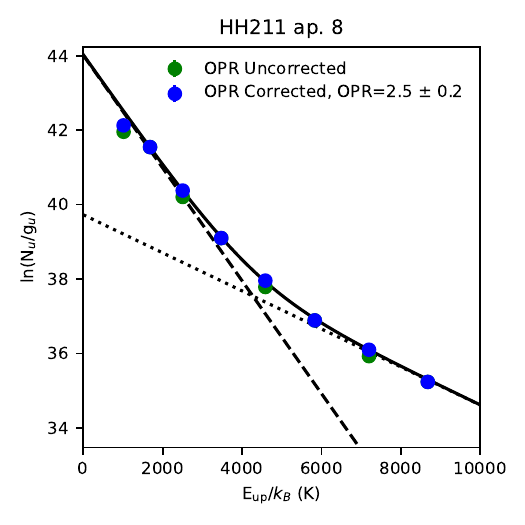}
    \caption{Example of the effect of correction for ortho-to-para H$_2$ ratio for IRAS 23385+6053 aperture 1 and HH 211  aperture 8. The effect on the H$_2$ rotation diagrams is shown before (green circles) and after (blue circles). The best-fit to the corrected data for the warm and hot H$_2$ components are overlaid as dashed and dotted lines, respectively.}
    \label{fig:opr_corr_example}
\end{figure}

\begin{figure}[htb]
    \centering
    \includegraphics[width=0.4\textwidth]{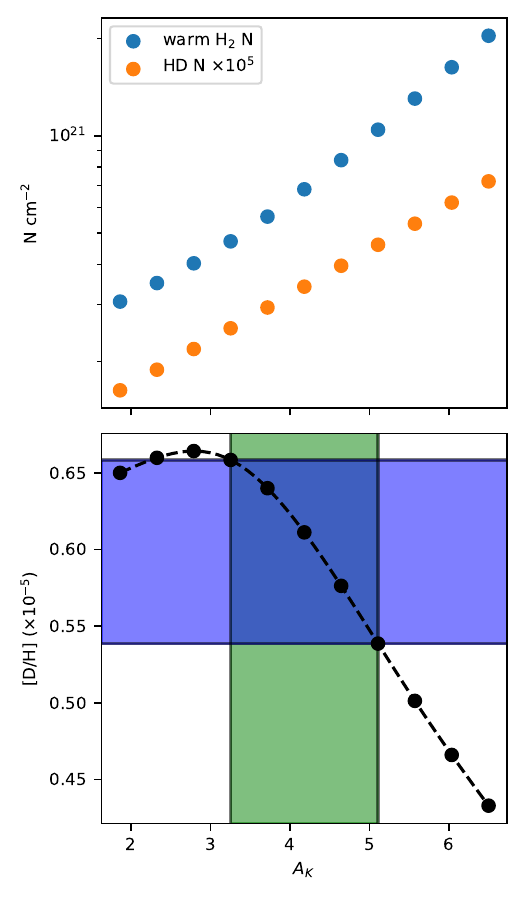}
    \caption{Example of the effect of extinction uncertainty on the derived column densities and [D/H] for IRAS 23385+6053 ap 1. {\it Upper panel}: Effect of varying extinction on derived column density for the warm H$_2$ component and HD. {\it Lower panel}: Effect of varying extinction on the derived $\mathrm{[D/H]}$. The uncertainty in $A_K$ from the single component fit to S(1)-S(4) transitions of H$_2$ is shown as the green shaded region, while the equivalent uncertainty in $\mathrm{[D/H]}$ is shown as the blue shaded region.}
    \label{fig:extinction_uncertainty}
\end{figure}

The KP5 \citep{Pontoppidan2024} extinction curve in the MIRI/MRS wavelength range used to derive $A_K$ from the H$_2$ S(1) to S(4) lines is shown in Fig. \ref{fig:ext_curve_and_lines}. We show an example of deriving the extinction correction for IRAS 23385+6053 aperture 1 and HH 211  aperture 8 in Fig. \ref{fig:ext_corr_example}, and an example of the correction for the ortho-to-para H$_2$ ratio in Figure \ref{fig:opr_corr_example}. The effect of extinction on the uncertainty in our warm H$_2$ and HD column column densities and derived [D/H] are shown in Fig. \ref{fig:extinction_uncertainty}.

\clearpage

\FloatBarrier

\clearpage
\section{Rotation Diagram Fit Results}
\label{app:rot_diagram_results}

In table \ref{tab:rot_diagram_results}, we provide the results of the rotation diagram fits for H$_2$ and HD to each aperture described in Section \ref{ssec:rotation_diagrams}.

\begin{table}[htb]
    \tiny
    \caption{Rotation Diagram Fit Results.}
\begin{tabular}{lllllllll}
\toprule\toprule
               Source & H$_2$ $N_\mathrm{warm}$ & H$_2$ $T_\mathrm{warm}$ & H$_2$ $N_\mathrm{hot}$ & H$_2$ $T_\mathrm{hot}$ &       H$_2$ OPR &               HD $N$ &          HD $T$ &         $A_K$ \\
\midrule
G28-IRS2 ap. 1 & $3.05 \pm 1.36$ (20) & $455 \pm 52$ & $4.68 \pm 24.34$ (17) & $3000^*$ & $2.52 \pm 0.77$ & $< 1.20 (16)$ & $455^\dagger$ & $5.7 \pm 0.1$ \\
G28-P1-A ap. 1 & $6.16 \pm 3.13$ (20) & $466 \pm 65$ & $1.53 \pm 2.81$ (18) & $3000^*$ & $1.66 \pm 0.68$ & $< 9.55 (15)$ & $466^\dagger$ & $6.0 \pm 0.9$ \\
G31-A ap. 1 & $2.17 \pm 1.24$ (22) & $395 \pm 59$ & $8.19 \pm 23.30$ (18) & $3000^*$ & $1.02 \pm 0.59$ & $< 6.53 (17)$ & $395^\dagger$ & $9.3 \pm 3.0$ \\
IRAS 18089-1732 ap. 1 & $1.52 \pm 1.03$ (22) & $401 \pm 52$ & $9.51 \pm 4.67$ (18) & $3000^*$ & $2.67 \pm 0.59$ & $< 8.88 (16)$ & $401^\dagger$ & $8.8 \pm 2.4$ \\
IRAS 23385+6053 ap. 1 & $6.83 \pm 0.99$ (20) & $552 \pm 38$ & $3.61 \pm 5.13$ (18) & $2653 \pm 2310$ & $1.76 \pm 0.17$ & $3.41 \pm 1.12$ (15) & $711 \pm 59$  & $4.2 \pm 0.9$ \\
BHR71-IRS1 ap. 1 & $2.17 \pm 0.59$ (20) & $853 \pm 63$ & $3.19 \pm 15.63$ (18) & $3000^*$ & $2.67 \pm 0.67$ & $3.36 \pm 0.85$ (15) & $849 \pm 63$  & $3.2 \pm 0.7$ \\
BHR71-IRS1 ap. 2 & $5.99 \pm 0.85$ (19) & $851 \pm 65$ & $2.56 \pm 1.63$ (18) & $2747 \pm 982$ & $3.00 \pm 0.25$ & $5.15 \pm 1.41$ (14) & $1043 \pm 98$  & $1.7 \pm 0.4$ \\
BHR71-IRS1 ap. 3 & $3.66 \pm 0.18$ (19) & $900 \pm 35$ & $1.90 \pm 0.82$ (18) & $3000^*$ & $2.90 \pm 0.14$ & $3.76 \pm 0.57$ (14) & $980 \pm 41$  & $1.1 \pm 0.0$ \\
BHR71-IRS1 ap. 4 & $2.16 \pm 0.10$ (19) & $659 \pm 11$ & $2.75 \pm 0.87$ (17) & $3000^*$ & $3.00 \pm 0.12$ & $< 4.34 (14)$ & $659^\dagger$ & $1.2 \pm 0.1$ \\
HH 211 ap. 1 & $1.52 \pm 0.19$ (20) & $654 \pm 28$ & $6.13 \pm 1.87$ (18) & $1463 \pm 93$ & $2.99 \pm 0.09$ & $1.62 \pm 0.40$ (15) & $649 \pm 40$  & $1.3 \pm 0.1$ \\
HH 211 ap. 2 & $1.48 \pm 0.05$ (20) & $560 \pm 20$ & $7.72 \pm 3.04$ (18) & $1160 \pm 89$ & $3.00 \pm 0.13$ & $2.09 \pm 0.06$ (15) & $518 \pm 4$  & $1.2 \pm 0.1$ \\
HH 211 ap. 3 & $1.41 \pm 0.05$ (20) & $619 \pm 28$ & $3.95 \pm 4.06$ (18) & $1392 \pm 288$ & $3.00 \pm 0.14$ & $2.04 \pm 0.45$ (15) & $542 \pm 27$  & $1.2 \pm 0.2$ \\
HH 211 ap. 4 & $6.47 \pm 0.31$ (19) & $572 \pm 13$ & $9.24 \pm 2.82$ (17) & $1551 \pm 130$ & $2.91 \pm 0.08$ & $< 6.19 (14)$ & $572^\dagger$ & $1.0 \pm 0.2$ \\
HH 211 ap. 5 & $7.40 \pm 1.43$ (19) & $623 \pm 38$ & $4.30 \pm 6.80$ (17) & $2498 \pm 1904$ & $2.35 \pm 0.15$ & $< 5.17 (14)$ & $623^\dagger$ & $1.4 \pm 1.1$ \\
HH 211 ap. 6 & $4.75 \pm 0.25$ (19) & $552 \pm 17$ & $5.87 \pm 3.74$ (17) & $1627 \pm 342$ & $2.76 \pm 0.07$ & $< 1.13 (15)$ & $552^\dagger$ & $0.9 \pm 0.4$ \\
HH 211 ap. 7 & $1.75 \pm 0.31$ (21) & $529 \pm 51$ & $3.22 \pm 3.07$ (19) & $1696 \pm 369$ & $1.95 \pm 0.37$ & $3.62 \pm 1.19$ (15) & $885 \pm 67$  & $1.7 \pm 2.1$ \\
HH 211 ap. 8 & $2.13 \pm 0.18$ (20) & $660 \pm 50$ & $8.97 \pm 6.02$ (18) & $1956 \pm 441$ & $2.53 \pm 0.21$ & $5.71 \pm 1.76$ (14) & $1271 \pm 146$  & $1.1 \pm 0.8$ \\
HH 211 ap. 9 & $2.56 \pm 0.54$ (20) & $639 \pm 49$ & $9.47 \pm 4.96$ (18) & $2215 \pm 620$ & $2.08 \pm 0.24$ & $9.17 \pm 3.33$ (14) & $1198 \pm 132$  & $1.5 \pm 1.4$ \\
IRAS 4B ap. 1 & $3.12 \pm 0.66$ (21) & $565 \pm 43$ & $3.48 \pm 2.35$ (19) & $2249 \pm 725$ & $2.13 \pm 0.18$ & $3.05 \pm 1.13$ (15) & $1690 \pm 376$  & $4.1 \pm 1.5$ \\
IRAS 4B ap. 2 & $3.79 \pm 1.00$ (21) & $552 \pm 52$ & $2.47 \pm 1.67$ (19) & $2679 \pm 1261$ & $2.13 \pm 0.26$ & $3.84 \pm 0.98$ (15) & $1694 \pm 243$  & $3.8 \pm 1.5$ \\
L1448-mm ap. 1 & $1.70 \pm 0.16$ (20) & $730 \pm 33$ & $2.08 \pm 1.52$ (18) & $2620 \pm 1181$ & $3.00 \pm 0.20$ & $2.30 \pm 0.30$ (15) & $811 \pm 26$  & $1.7 \pm 0.2$ \\
L1448-mm ap. 2 & $1.01 \pm 0.05$ (20) & $640 \pm 12$ & $8.74 \pm 9.19$ (17) & $1891 \pm 667$ & $3.00 \pm 0.14$ & $1.87 \pm 0.07$ (15) & $472 \pm 3$  & $1.3 \pm 0.2$ \\
L1448-mm ap. 3 & $1.04 \pm 0.03$ (20) & $649 \pm 8$ & $1.01 \pm 0.42$ (18) & $1967 \pm 334$ & $2.98 \pm 0.09$ & $1.90 \pm 0.24$ (15) & $481 \pm 17$  & $1.1 \pm 0.0$ \\
L1448-mm ap. 4 & $4.09 \pm 0.15$ (19) & $611 \pm 14$ & $6.46 \pm 8.93$ (16) & $3000^*$ & $3.00 \pm 0.08$ & $< 4.61 (14)$ & $611^\dagger$ & $1.3 \pm 0.0$ \\
Ser-emb 8 (N) ap. 1 & $4.01 \pm 1.11$ (20) & $747 \pm 213$ & $2.16 \pm 212.38$ (19) & $1519 \pm 1826$ & $1.39 \pm 0.34$ & $3.01 \pm 0.53$ (15) & $749 \pm 36$  & $3.5 \pm 1.9$ \\
Ser-emb 8 (N) ap. 2 & $5.35 \pm 1.10$ (20) & $696 \pm 91$ & $4.14 \pm 27.14$ (18) & $3000^*$ & $1.33 \pm 0.23$ & $2.50 \pm 1.00$ (15) & $822 \pm 100$  & $3.2 \pm 1.6$ \\
\midrule
\end{tabular}
    \\ Numbers formatted as $a \pm b$ $(c)$ indicate $(a \pm b) \times 10^c$. Error bars on column density and rotational temperature do not include uncertainty from extinction. \\
    $^*$Temperature of hot component reaches the fit limit of $3000$ K and is highly uncertain. \\
    $^\dagger$Temperature of HD assumed to be the same as warm H$_2$ for column density upper limit determination.
    
    \label{tab:rot_diagram_results}
\end{table}

\end{appendix}
\end{document}